%
%
%
%
%
%


\magnification=1200
\hsize=31pc
\vsize=55 truepc
\hfuzz=2pt
\vfuzz=4pt
\pretolerance=5000
\tolerance=5000
\parskip=0pt plus 1pt
\parindent=16pt
%

%
%
\font\fourteenrm=cmr10 scaled \magstep2
\font\fourteeni=cmmi10 scaled \magstep2
\font\fourteenbf=cmbx10 scaled \magstep2
\font\fourteenit=cmti10 scaled \magstep2
\font\fourteensy=cmsy10 scaled \magstep2

%
\font\large=cmbx10 scaled \magstep1

%

%

%

%
\font\eightrm=cmr8  
\font\eighti=cmmi8
\font\eightbf=cmbx8
\font\eightit=cmti8

\font\eightsy=cmsy8
\font\sixrm=cmr6
\font\sixi=cmmi6
\font\sixsy=cmsy6

%
\def\tenpoint{\def\rm{\fam0\tenrm}%
  \textfont0=\tenrm \scriptfont0=\sevenrm 
                      \scriptscriptfont0=\fiverm
  \textfont1=\teni  \scriptfont1=\seveni 
                      \scriptscriptfont1=\fivei
  \textfont2=\tensy \scriptfont2=\sevensy 
                      \scriptscriptfont2=\fivesy
  \textfont3=\tenex   \scriptfont3=\tenex 
                      \scriptscriptfont3=\tenex
  \textfont\itfam=\tenit  \def\it{\fam\itfam\tenit}%
  \textfont\slfam=\tensl  \def\sl{\fam\slfam\tensl}%
  \textfont\bffam=\tenbf  \scriptfont\bffam=\sevenbf
                            \scriptscriptfont\bffam=\fivebf
                            \def\bf{\fam\bffam\tenbf}%
  \normalbaselineskip=20 truept
  \setbox\strutbox=\hbox{\vrule height14pt depth6pt
width0pt}%
  \let\sc=\eightrm \normalbaselines\rm}
\def\eightpoint{\def\rm{\fam0\eightrm}%
  \textfont0=\eightrm \scriptfont0=\sixrm 
                      \scriptscriptfont0=\fiverm
  \textfont1=\eighti  \scriptfont1=\sixi
                      \scriptscriptfont1=\fivei
  \textfont2=\eightsy \scriptfont2=\sixsy
                      \scriptscriptfont2=\fivesy
  \textfont3=\tenex   \scriptfont3=\tenex
                      \scriptscriptfont3=\tenex
  \textfont\itfam=\eightit  \def\it{\fam\itfam\eightit}%
  \textfont\bffam=\eightbf  \def\bf{\fam\bffam\eightbf}%
  \normalbaselineskip=16 truept
  \setbox\strutbox=\hbox{\vrule height11pt depth5pt width0pt}}
\def\fourteenpoint{\def\rm{\fam0\fourteenrm}%
  \textfont0=\fourteenrm \scriptfont0=\tenrm 
                      \scriptscriptfont0=\eightrm
  \textfont1=\fourteeni  \scriptfont1=\teni 
                      \scriptscriptfont1=\eighti
  \textfont2=\fourteensy \scriptfont2=\tensy 
                      \scriptscriptfont2=\eightsy
  \textfont3=\tenex   \scriptfont3=\tenex 
                      \scriptscriptfont3=\tenex
  \textfont\itfam=\fourteenit  \def\it{\fam\itfam\fourteenit}%
  \textfont\bffam=\fourteenbf  \scriptfont\bffam=\tenbf
                             \scriptscriptfont\bffam=\eightbf
                             \def\bf{\fam\bffam\fourteenbf}%
  \normalbaselineskip=24 truept
  \setbox\strutbox=\hbox{\vrule height17pt depth7pt width0pt}%
  \let\sc=\tenrm \normalbaselines\rm}
\def\today{\number\day\ \ifcase\month\or
  January\or February\or March\or April\or May\or June\or
  July\or August\or September\or October\or November\or
December\fi
  \space \number\year}

%
\newcount\secno      
\newcount\subno      
\newcount\subsubno   
\newcount\appno      
\newcount\tableno    
\newcount\figureno   
%

%
\normalbaselineskip=20 truept
\baselineskip=20 truept

%
%
\def\title#1
   {\vglue1truein
   {\baselineskip=24 truept
    \pretolerance=10000
    \raggedright
    \noindent \fourteenpoint\bf #1\par}
    \vskip1truein minus36pt}
%

%
\def\author#1
  {{\pretolerance=10000
    \raggedright
    \noindent {\large #1}\par}}

%
\def\address#1
   {\bigskip
    \noindent \rm #1\par}

%
\def\shorttitle#1
   {\vfill
    \noindent \rm Short title: {\sl #1}\par
    \medskip}

%
\def\pacs#1
   {\noindent \rm PACS number(s): #1\par
    \medskip}

%
\def\jnl#1
   {\noindent \rm Submitted to: {\sl #1}\par
    \medskip}

%
\def\date
   {\noindent Date: \today\par
    \medskip}

%
\def\beginabstract
   {\vfill\eject
    \noindent {\bf Abstract. }\rm}

%
\def\keyword#1
   {\bigskip
    \noindent {\bf Keyword abstract: }\rm#1}

%
\def\endabstract
   {\par
    \vfill\eject}

%
%

%
\def\entry#1#2#3
   {\noindent
    \hangindent=20pt
    \hangafter=1
    \hbox to20pt{#1 \hss}#2\hfill #3\par}

%
\def\subentry#1#2#3
   {\noindent
    \hangindent=40pt
    \hangafter=1
    \hskip20pt\hbox to20pt{#1 \hss}#2\hfill #3\par}
\def\checkforsub{\futurelet\nexttok\decide}
\def\ssf{\relax}
\def\decide{\if\nexttok\ssf\let\endspace=\nospace
                \else\let\endspace=\extraspace\fi\endspace}
\def\nospace{\nobreak\par\nobreak}
%
%
\def\section#1{%
    \goodbreak
    \vskip50pt plus12pt minus12pt
    \nobreak
    \gdef\extraspace{\nobreak\bigskip\noindent\ignorespaces}%
    \noindent
    \subno=0 \subsubno=0 
    \global\advance\secno by 1
    \noindent {\bf \the\secno. #1}\par\checkforsub}

%
\def\subsection#1{%
     \goodbreak
     \vskip24pt plus12pt minus6pt
     \nobreak
     \gdef\extraspace{\nobreak\medskip\noindent\ignorespaces}%
     \noindent
     \subsubno=0
     \global\advance\subno by 1
     \noindent {\sl \the\secno.\the\subno. #1\par}\checkforsub}

%
\def\subsubsection#1{%
     \goodbreak
     \vskip20pt plus6pt minus6pt
     \nobreak\noindent
     \global\advance\subsubno by 1
     \noindent {\sl \the\secno.\the\subno.\the\subsubno. #1}\null. 
     \ignorespaces}

%
\def\appendix#1
   {\vskip0pt plus.1\vsize\penalty-250
    \vskip0pt plus-.1\vsize\vskip24pt plus12pt minus6pt
    \subno=0
    \global\advance\appno by 1
    \noindent {\bf Appendix \the\appno. #1\par}
    \bigskip
    \noindent}

%
\def\subappendix#1
   {\vskip-\lastskip
    \vskip36pt plus12pt minus12pt
    \bigbreak
    \global\advance\subno by 1
    \noindent {\sl \the\appno.\the\subno. #1\par}
    \nobreak
    \medskip
    \noindent}

%


%

%
\def\tabcaption#1
   {\global\advance\tableno by 1
    \noindent {\bf Table \the\tableno.} \rm#1\par
    \bigskip}

%

%

%

%

%

%
\def\figcaption#1
   {\global\advance\figureno by 1
    \noindent {\bf Figure \the\figureno.} \rm#1\par
    \bigskip}

%
\def\references
     {\vfill\eject 
     {\noindent \bf References\par}
      \parindent=0pt
      \bigskip}

%

%
\def\refjl#1#2#3#4
   {\hangindent=16pt
    \hangafter=1
    \rm #1
   {\frenchspacing\sl #2
    \bf #3}
    #4\par}

%
\def\refbk#1#2#3
   {\hangindent=16pt
    \hangafter=1
    \rm #1
   {\frenchspacing\sl #2}
    #3\par}

%
\def\numrefjl#1#2#3#4#5
   {\parindent=40pt
    \hang
    \noindent
    \rm {\hbox to 30truept{\hss #1\quad}}#2
   {\frenchspacing\sl #3\/
    \bf #4}
    #5\par\parindent=16pt}

%
\def\numrefbk#1#2#3#4
   {\parindent=40pt
    \hang
    \noindent
    \rm {\hbox to 30truept{\hss #1\quad}}#2
   {\frenchspacing\sl #3\/}
    #4\par\parindent=16pt}

%

\def\ref#1{\noindent \hbox to 21pt{\hss 
#1\quad}\frenchspacing\ignorespaces}

%
\def\frac#1#2{{#1 \over #2}}

%

%

%


\chardef\ii="10

%

%

%

\catcode`\@=11
\def\vfootnote#1{\insert\footins\bgroup
    \interlinepenalty=\interfootnotelinepenalty
    \splittopskip=\ht\strutbox 
    \splitmaxdepth=\dp\strutbox \floatingpenalty=20000
    \leftskip=0pt \rightskip=0pt \spaceskip=0pt \xspaceskip=0pt
    \noindent\eightpoint\rm #1\ \ignorespaces\footstrut\futurelet\next\fo@t}

%
%
\def\eq(#1){\hfill\llap{(#1)}}
\catcode`\@=12
%
%



%
%
\def\CQG{Classical Quantum Grav.}





%
%

%
%

%
%

%

%

%

%
\def\gap{\;\lower3pt\hbox{$\buildrel > \over \sim$}\;}
%
%
\def\lap{\;\lower3pt\hbox{$\buildrel < \over \sim$}\;}
\def\tqs{\hbox to 25pt{\hfil}}


%
%
%
\def\LaTeX{L\kern-.26em \raise.6ex\hbox{\fiverm A}%
   \kern-.15em\TeX}%
\def\AmSTeX{%
{$\cal{A}$}\kern-.1667em\lower.5ex\hbox{%
 $\cal{M}$}\kern-.125em{$\cal{S}$}-\TeX}



\def\square{\hbox{\rlap{$\sqcap$}$\sqcup$}}

\title{Quadrupole-quadrupole gravitational waves}

\author{Luc Blanchet}
\address{D\'epartement d'Astrophysique Relativiste et de Cosmologie
(UPR 176 du CNRS), Observatoire de Paris, 92195 Meudon Cedex, France}

\shorttitle{Quadrupole-quadrupole gravitational waves}
\pacs{04.25.Nx, 04.30.Db}
\jnl{\CQG}
\date

\beginabstract
This paper investigates the non-linear self-interaction of quadrupole
gravitational waves generated by an isolated system. The vacuum
Einstein field equations are integrated in the region exterior to the
system by means of a post-Minkowskian algorithm. Specializing in the
quadrupole-quadrupole interaction (at the quadratic non-linear order),
we recover the known results concerning the non-local modification of  
the ADM mass-energy of the system accounting for the emission of quadrupole  
waves, and the non-local memory effect due to the re-radiation by the  
stress-energy distribution of linear waves. Then we compute all the local  
(instantaneous) terms which are associated in the quadrupole-quadrupole
metric with the latter non-local effects.  Expanding the metric at large
distances from the system, we obtain the corresponding radiation-field
observables, including all non-local and transient contributions.  This
permits notably the completion of the observable quadrupole moment at
the 5/2 post-Newtonian order.
\endabstract

\section{Introduction}

In general relativity, the multipole moments of any finite distribution
of energy and momentum interact with each other in vacuum, through the
non-linearities of the field equations. In particular, the multipole
moments which describe the gravitational waves emitted by an isolated system
do not evolve independently, but rather couple together (including with  
themselves), giving rise to non-linear physical effects.

The simplest multipole interaction which contributes to the radiation
field is that between the (mass-type) quadrupole moment $M_{ij}$ and the
mass monopole $M$. The latter moment is the constant mass-energy of the
source as measured at spatial infinity (ADM mass). Associated with the
multipole interaction $M_{ij}\times M$ is the non-linear effect of
tails. This effect is due to the backscatter of linear waves (described
by $M_{ij}$) onto the space-time curvature generated by the mass-energy  
$M$.  The tails can be computed within the theory of gravitational
perturbations of the Schwarzschild background (see e.g.  [1--3]).  A
consequence of the existence of tails is the non-locality in time, as
the tails are in the form of integrals depending on the history of the
source from $-\infty$ in the past to the retarded time $t-r/c$ [4--7].
It is known that the tails appear both in the radiation field and in the
radiation reaction forces at the 3/2 post-Newtonian order (1.5PN, or
order $c^{-3}$ when $c \to \infty$) relatively to the quadrupole
radiation [8,9].

Next in complexity is the interaction of the quadrupole moment with
itself, or quadrupole self-interaction $M_{ij}\times M_{kl}$. Two
closely related non-local (or hereditary) effects are known with this
particular interaction. As shown by Bonnor and collaborators [10,11,4,6],  
the total mass $M$ gets modified by a non-local integral accounting for  
the energy which is radiated by the quadrupole waves (there is agreement  
with the Einstein quadrupole formula). On the other hand the radiation field  
involves a non-local contribution whose physical origin is the re-emission  
of waves by the linear waves [12--16,9]. This contribution can be easily
computed by using as the source of waves in the right side of the Einstein
field equations the effective stress-energy tensor of gravitational waves  
(averaged over several wavelengths). As shown by Christodoulou [14] and  
Thorne [15] this implies a permanent change in the wave amplitude  
from before to afterward a burst of gravitational waves, which can  
be interpreted as the contribution of gravitons in the known formulas  
for the linear memory [17,18]. The latter non-linear memory integral
appears at the 2.5PN order in the radiation field (1PN order relatively  
to the tail integral).

Now the metric which corresponds to the quadrupole-quadrupole interaction  
involves also, besides the non-local contributions, many terms which
by contrast depend on the multipole moments at the sole retarded instant  
$t-r/c$. In the following we shall often qualify these local terms as  
instantaneous (following the terminology of [8]). The instantaneous terms  
in the radiation field (to order $1/r$) are transient in the sense that they  
return to zero after the passage of a burst of gravitational waves.  
On physical grounds, it can be argued that the instantaneous terms do not  
play a very important role. However, these terms do exist and form an integral  
part of the field generated by very relativistic sources like inspiraling  
compact binaries (see e.g. [19]). In fact, the complete quadrupole-quadrupole
metric including all the hereditary and instantaneous contributions will be  
needed in the construction of very accurate theoretical wave forms to be  
used by the future detectors LIGO and VIRGO.

The present paper is devoted to the computation of the quadrupole-quadrupole
metric, using the so-called multipolar-post-Minkowskian method proposed  
by Blanchet and Damour [20,21] after previous work by Bonnor [10]
(his double-series approximation method) and Thorne [22] (how to start the  
post-Minkowskian iteration using STF multipole moments). The instantaneous  
quadrupole-quadrupole terms have never been computed within the
multipolar-post-Minkowskian method. However, they have been computed by Hunter
and Rotenberg [6] within the double-series method, in the case where the
source is axi-symmetric.  The non-linear interaction between quadrupoles
has also received attention more recently within the double-series
method [23].

The main result of this paper (having in mind the application to astrophysical sources to be detected by
VIRGO and LIGO) is the completion of the observable quadrupole moment  
of a general isolated source at the 2.5PN order. To reach this result we also
take into account previous results concerning the tails at the 1.5PN order  
[9], and the multipole moments given by explicit integrals over the source to  
2.5PN order [24,25]. (Note that the 2.5PN approximation in the observable  
quadrupole moment gives no contribution to the phase evolution of inspiraling  
compact binaries [25], however it will be required when we compute  
the 2.5PN wave form.)

In a following paper [26], that we shall refer to as paper II, we shall
investigate the monopole-monopole-quadrupole interaction (at the cubic
non-linear order) which enters the radiation field at the 3PN
approximation.  The present paper and paper II are part of the program
of computing the field generated by inspiraling compact binaries with
3PN and even 3.5PN accuracy (see [27--30] for why such a very high
accuracy is necessary).

The plan of the paper is as follows. In Section 2 we summarize from
[20,21] the method for computing the field non-linearities. In Section
3 we investigate (following [8]) the general structure of the quadratic
non-linearities. Section 4 deals with the explicit computation of the
quadrupole-quadrupole metric. The results are presented in the form of
tables of numerical coefficients. Finally, in Section 5, we expand the
metric at infinity and obtain the observable moments in the radiation
field (essentially the quadrupole). The technical formulas for integrating the  
wave equation are relegated to Appendix A, and some needed results concerning  
the dipole-quadrupole interaction are derived in Appendix B. Henceforth we pose
$c=1$, except when we discuss the post-Newtonian order  
at the end of Section 5.

\section{Non-linearities in the external field} 

We summarize the method set up in [20] for the computation of the quadratic  
and higher non-linearities in the field generated by an isolated system.  
The computation is performed in the exterior (vacuum) weak-field region of the  
system, where the components of the gravitational field $h^{\mu\nu}$
are numerically small as compared to one. Here $h^{\mu\nu}$ denotes the metric
deviation $h^{\mu\nu}=\sqrt{-g} g^{\mu\nu} -\eta^{\mu\nu}$, with $g^{\mu\nu}$  
the inverse and $g$ the determinant of the metric $g_{\mu\nu}$, and  
$\eta^{\mu\nu}$ the Minkowski metric $\rm diag (-1,1,1,1)$. The field
$h^{\mu\nu}$ admits in the exterior weak-field region a post-Minkowskian
expansion,

$$ h^{\mu\nu} = G h^{\mu\nu}_1 + G^2 h^{\mu\nu}_2 + ... + G^n
  h^{\mu\nu}_n + ... \ , \eqno(2.1) $$

\noindent where $G$ is the Newtonian gravitational constant, which plays
here the role of book-keeping parameter in the non-linearity expansion.
The first term in (2.1) satisfies the vacuum Einstein equations
linearized around the Minkowski metric. In harmonic (or De Donder)
coordinates this means two equations,

$$ \eqalignno{
    \square h^{\mu\nu}_1 &=0 \ , &(2.2a) \cr
     \partial_\nu h^{\mu\nu}_1 &=0\ . &(2.2b)  \cr }
$$
In the first equation $\square$ denotes the flat space-time wave
(d'Alembertian) operator. The second equation (divergenceless of the
field) is the harmonic gauge condition. The equations (2.2), supplemented  
by the condition of retarded potentials, are solved by means of a multipolar  
expansion. It is known that only two sets of multipole moments, the mass-type  
moments $M_L$ and current-type moments $S_L$ (both depending on the retarded  
time $t-r$), are sufficient to parametrize the general multipole expansion  
(see e.g.  [22]). In terms of these moments the general solution reads [22]

$$ \eqalignno{
 h^{00}_1 =& - 4
  \sum_{\ell \geq 0} {(-)^{\ell} \over \ell !} \partial_ L
   \left[r^{-1}M_L (t- r) \right] \ ,  &(2.3a)\cr
 h^{0i}_1 =& 4 \sum_{\ell \geq 1}
   {(-)^{\ell} \over \ell !} \partial_{L-1}
   \left[r^{-1} M^{(1)}_{iL-1}
    (t- r) \right] \cr
  &+ 4 \sum_{\ell \geq 1}
   {(-)^{\ell} \ell \over (\ell +1)!} \varepsilon_{iab}\partial_{aL-1}
  \left[r^{-1}S_{bL-1} (t- r) \right] \ , &(2.3b) \cr
  h^{ij}_1 =& - 4
   \sum_{\ell \geq 2} {(-)^{\ell} \over \ell !} \partial_{L-2}
   \left[r^{-1} M^{(2)}_{ijL-2}
   (t- r) \right]  \cr
  &- 8 \sum_{\ell \geq 2} {(-)^{\ell} \ell \over (\ell +1)!}
   \partial_{aL-2} \left[r^{-1}\varepsilon_{ab(i}   S^{(1)}_{\!j)bL-2}
   (t- r) \right]\ . &(2.3c) \cr } $$
Here the superscript $(n)$ denotes $n$ time derivatives. The index $L$
is a shorthand for a multi-index composed of $\ell$ indices,
$L=i_1i_2...i_\ell$ (similarly, $L-1=i_1i_2...i_{\ell-1}$,
$aL-2=ai_1...i_{\ell-2}$, and so on), and $\partial_L$ denotes a product
of $\ell$ space derivatives, $\partial_L=\partial_{i_1} \partial_{i_2}
...\partial_{i_\ell}$. The multipole moments $M_L$ and $S_L$ are
symmetric and tracefree (STF) with respect to all their indices (see
[31] for a recapitulation of our notation and conventions).

{\it A priori} the multipole moments $M_L(t)$ and $S_L(t)$ are arbitrary
functions of time. They describe potentially the physics of a general
isolated source as seen in its exterior field [22,20]. The only
physical restriction is that the mass monopole $M$ (total mass-energy  
or ADM mass), the mass dipole $M_i$ (position of the center of mass times  
the mass), and the current dipole $S_i$ (total angular momentum) are constant.  
Technically speaking, this is a consequence of the harmonic gauge condition  
(2.2b) (see e.g.  [22]). In this paper we shall set the mass dipole $M_i$  
to zero by shifting the origin of coordinates to the center of mass.  
In order to describe an isolated system, we must implement a condition of
no-incoming radiation, ensuring that the radiation field is entirely
generated by the system. We assume that the field is stationary in the
remote past, i.e.  that the moments $M_L(t)$ and $S_L(t)$ are constant
before some finite instant in the past, say when $t \leq - {\cal T}$ 
(hence $M_i$ cannot be a linear function of time and is necessarily 
constant or zero). This
assumption may seem to be somewhat restrictive, but we can check {\it a
posteriori} that the formulas derived in this paper and paper II admit a
well-defined limit when $-{\cal T} \rightarrow -\infty$ in more general
physical situations, such as the formation of the system by initial
gravitational scattering.  The multipole moments $M_L(t)$ and $S_L(t)$,
subject to the previous restrictions, play the role of ``seed'' moments
for the construction of the exterior field (2.1).  In particular we
shall express the results of this paper in terms of products
of $M_{ij}$ with itself. We shall not use the expressions of the multipole
moments $M_L$ and $S_L$ as explicit integrals over the source. However,  
these expressions are known in the post-Newtonian approximation [24,25], 
and should be used in applications.

The coefficient of $G^2$ in (2.1) is the quadratically non-linear
metric, whose precise definition we recall. The field equations for this  
coefficient read, still using harmonic coordinates,

$$ \eqalignno{
 \square h^{\mu\nu}_2 &= N^{\mu\nu}_2 \ , &(2.4a) \cr
 \partial_\nu h^{\mu\nu}_2 &= 0 \ . &(2.4b) \cr } $$
The d'Alembertian equation involves a quadratic source
$N^{\mu \nu}_2=N^{\mu \nu}(h_1, h_1)$ generated by the linearized
gravitational field (2.3), where

$$ \eqalignno{
 N^{\mu\nu} (h,h) =&- h^{\rho\sigma} \partial_\rho \partial_\sigma
 h^{\mu\nu} + {1\over 2} \partial^\mu h_{\rho\sigma} \partial^\nu
 h^{\rho\sigma} - {1\over 4} \partial^\mu h \partial^\nu h \cr
&-2 \partial^{(\mu} h_{\rho\sigma} \partial^\rho h^{\nu)\sigma}
  +\partial_\sigma h^{\mu\rho} (\partial^\sigma h^\nu_\rho + \partial_\rho
  h^{\nu\sigma}) \cr
&+ \eta^{\mu\nu} \left[ -{1\over 4}\partial_\lambda h_{\rho\sigma}
  \partial^\lambda h^{\rho\sigma} +{1\over 8}\partial_\rho h \partial^\rho h
  +{1\over 2}\partial_\rho h_{\sigma\lambda} \partial^\sigma
  h^{\rho\lambda}\right]\ .
                 & (2.5) \cr }
$$
{}From (2.4b) we deduce

$$ \partial_\nu N^{\mu\nu}_2 = 0 \ . \eqno(2.6) $$
Then the quadratic metric $h^{\mu \nu}_2$, solving (2.4) and the
condition of stationarity in the past, is obtained as the sum of two
distinct contributions,

$$  h^{\mu\nu}_2 = u^{\mu\nu}_2 + v^{\mu\nu}_2 \ . \eqno(2.7) $$

Basically the first contribution $u^{\mu \nu}_2$ is the retarded
integral of the source $N^{\mu \nu}_2$. However, in our case the source
is in the form of a multipole expansion (valid only in the exterior of
the system and singular at $r=0$), so we cannot apply directly the usual
retarded integral operator, whose range of integration intersects the
system at retarded time. A way out of this problem, proposed in [20],
consists of multiplying the actual source term $N^{\mu \nu}_2$ by a
factor $(r/r_0)^B$, where $B$ is a complex number and $r_0$ denotes a
certain constant having the dimension of a length. When the real part of
$B$ is large enough all the power-like (except for logarithms) singularities
of the multipole expansion at the spatial origin of the coordinates
$r=0$ are cancelled.  [Actually we consider separately each multipolar
pieces, with given multipolarities $\ell$, so that the maximal power of
the singularities is finite, and $B$ can indeed be chosen in such a
way.] Applying the retarded integral on each multipolar pieces of the
product $(r/r_0)^B N^{\mu \nu}_2$ results in a function of $B$ whose
definition can be analytically continued to a neighbourhood of $B=0$, at
which value it admits a Laurent expansion.  The finite part at $B=0$ (in
short ${\rm FP}_{B=0}$) of the latter expansion is our looked-for
solution, as it satisfies the correct wave equation ($\square
u^{\mu\nu}_2=N_2^{\mu\nu}$), and is like $N_2^{\mu\nu}$ in the form of a
multipole expansion. Note that the latter process represents simply a
convenient mean to find a solution of the wave equation whose source is
in the form of a multipole expansion.  Other processes could be used as
well, but this one is particularly powerful as it yields many explicit
formulas to be used in practical computations (see Appendix A below and
Appendix A of paper II).  Hence the first contribution in (2.7) reads

$$ u^{\mu\nu}_2 = {\rm FP}_{B=0}\, \square^{-1}_R \left[ \left( {r\over r_0}
   \right)^B N^{\mu\nu}_2 \right] \ , \eqno(2.8)  $$

\noindent where $\square ^{-1}_R$ denotes the usual retarded integral

$$ (\square^{-1}_R f)({\bf x},t) = -{1\over 4\pi} \int\!\!\!\int\!\!\!\int
   {d^3{\bf x}'\over |{\bf x}-{\bf x}'|} f ({\bf x}', t-|{\bf x}
   -{\bf x}'|)\ . \eqno(2.9) $$
When dealing with the metric at quadratic order, it can be proved that
the $B$-dependent retarded integral in (2.8) is actually finite when
$B\rightarrow 0$ (the finite part is not followed by any pole). So
$u^{\mu\nu}_2$ is simply given by the value at $B=0$ of the retarded
integral (see [8] and Section 4). But this is due to the
special structure of the quadratic source $N^{\mu\nu}_2$, and does not
remain true at cubic and higher non-linear approximations (see for
instance paper II).

The first contribution $u^{\mu\nu}_2$ solves (2.4a), but not the
harmonic gauge condition (2.4b). The divergence of $u^{\mu\nu}_2$, say
$w^{\mu}_2=\partial_{\nu}u^{\mu\nu}_2$, is {\it a priori} different
from zero. Using (2.6) we find

$$ w^\mu_2 = {\rm FP}_{B=0}\, \square^{-1}_R \left[ B \left( {r\over r_0}
        \right)^B {n_i\over r} N^{\mu i}_2 \right] \ . \eqno(2.10) $$
The explicit factor $B$ comes from the differentiation of $r^B$ in
(2.8) (we denote $n_i=\partial_i r=x^i/r$). Owing to this factor the
finite part in (2.10) is in fact a residue at $B=0$, or coefficient of
$B^{-1}$ in the Laurent expansion. [The source term in (2.10) has a
structure which is different from $N_2^{\mu\nu}$, and unlike in (2.8)
the integral admits in general a (simple) pole at $B=0$.] The second
contribution $v_2^{\mu\nu}$ in (2.7) is then defined in such a way as
to compensate exactly the ({\it a priori}) non-zero divergence $w_2^\mu$
of $u_2^{\mu\nu}$, while being a homogeneous solution of (2.4a). This is
possible because $w_2^\mu$ is a particular retarded solution of
$\square w_2^\mu=0$ (in the exterior region). As such, it admits a
unique multipolar decomposition in terms of four sets of STF tensors
$A_L$, $B_L$, $C_L$, $D_L$, namely

$$ \eqalignno{
w^0_2 &=\sum_{\ell\geq 0}\partial_L \left[r^{-1} A_L(t-r)\right]\ ,&(2.11a)\cr
w^i_2 &=\sum_{\ell\geq 0}\partial_{iL} \left[ r^{-1} B_L (t-r) \right] \cr
  &+ \sum_{\ell\geq 1} \left\{ \partial_{L-1} \left[ r^{-1} C_{iL-1} (t-r)
  \right] + \varepsilon_{iab} \partial_{aL-1} \left[r^{-1}
   D_{bL-1} (t-r) \right] \right\}\ .  &(2.11b) \cr  } $$
These tensors can be computed straigthforwardly from the known expression  
(2.10), and the contribution $v_2^{\mu\nu}$ is defined in terms of these  
tensors. A particular definition was proposed in the equations (4.13) of [20],  
where it was denoted $q_2^{\mu\nu}$. Here we shall define this second  
contribution slightly differently, and accordingly we use the different  
notation $v_2^{\mu\nu}$. The various components of $v^{\mu\nu}_2$ are
given by

$$ \eqalignno{
 v^{00}_2 =& - r^{-1} \hbox{$\int$} A + \partial_a \left[ r^{-1}
 \left(- \hbox{$\int$} A_a+ \hbox{$\int\!\!\int$} C_a -3B_a\right) \right]
    \ ,&(2.12a)\cr
  v^{0i}_2 =& ~r^{-1} \left( -\hbox{$\int$} C_i +3 B^{(1)}_i\right)
  - \varepsilon_{iab} \partial_a \left[ r^{-1} \hbox{$\int$} D_b \right] \cr
  &-\sum_{\ell\geq 2}\partial_{L-1}
 \left[ r^{-1} A_{iL-1} \right] \ , &(2.12b) \cr
 v^{ij}_2 =& - \delta_{ij} r^{-1} B + \sum_{\ell\geq 2} \biggl\{
   2 \delta_{ij}\partial_{L-1} \left[ r^{-1} B_{L-1}\right] -6
   \partial_{L-2(i} \left[ r^{-1} B_{j)L-2}\right] \cr
 &+ \partial_{L-2} \left[ r^{-1} (A^{(1)}_{ijL-2}
  + 3 B^{(2)}_{ijL-2} - C_{ijL-2}) \right]  \cr
 &  - 2 \partial_{aL-2}\left[ r^{-1} \varepsilon_{ab(i} D_{j)bL-2}
   \right] \biggr\}\ . &(2.12c) \cr   } $$
Like in (2.11) all the tensors are evaluated at the retarded time
$t-r$. We note that the formulas (2.12) are non-instantaneous, as they  
depend on the moments $M_L$ and $S_L$ at any time less than $t-r$  
through the first and second time anti-derivatives of $A$, $A_a$,  
$C_a$, denoted e.g.  by $\int A=\int_{-\infty}^{t-r} A (t')dt'$  
and $\int\!\!\int C_a= \int _{-\infty} ^{t-r} \int C_a(t')dt'$ (see [20]  
for discussions). [To quadratic order the tensors $A _L$, ..., $D_L$  
are given by some instantaneous functionals of the moments $M_L$ and  
$S_L$.] The main property of $v_2^{\mu \nu}$ is $\partial_\nu  
v_2^{\mu\nu}=-w_2^{\mu}$, which is easily checked on the expressions (2.12).  
Furthermore $\square v_2^{\mu \nu}=0$, so the quadratic metric (2.7) is,  
indeed, a solution of both the wave equation (2.4a) and the gauge  
condition (2.4b). Note that the spatial trace $v_2^{ii}=\delta^{ij}v_2^{ij}$  
is especially simple,

$$ v^{ii}_2 = -3 r^{-1} B\ . \eqno(2.12d) $$
The choice of definition (2.12) adopted here, which differs from the choice
adopted in [20], is for convenience in future work. Of course we are free
to adopt one definition or another because such a choice is equivalent 
to a choice of gauge. However, the definition (2.12) is slightly
preferable to the definition proposed in [20] when we want to express
the multipole moments $M_L$ and $S_L$ as integrals over the source.
Thus we take the present opportunity to redefine the construction of the
exterior metric using this new definition.  [Actually it can be checked
that all intermediate and final results of this paper and paper II are
independent of the choice of definition.]

To the third (cubic) and higher non-linear iterations the construction
of the external metric proceeds exactly along the same line, namely

$$  h^{\mu\nu}_n = u^{\mu\nu}_n + v^{\mu\nu}_n \ , \eqno(2.13) $$

\noindent where the first term $u_n^{\mu \nu}$ is the finite part of the
retarded integral of the source to the $n$th post-Minkowskian order,
$u_n^{\mu\nu}={\rm FP}_{B=0} \square _R^{-1}[(r/r_0)^B N_n^{\mu\nu} 
(h_1,...,h_{n-1})]$, and where the second term $v_n^{\mu\nu}$ is defined  
from the divergence $w_n^\mu =\partial_\nu u_n^{\mu\nu}$ by the same formulas  
(2.11)-(2.12). (See [20] for the proof that the construction of the metric  
can be implemented to any post-Minkowskian order.)

\section{Structure of the quadratically non-linear field}   

In this section we investigate the structure of the quadratic metric  
$h_2^{\mu \nu}$ defined by (2.7)-(2.12). For simplicity we omit most of the  
numerical coefficients and indices, so as to focus our attention on the  
basic structure of the metric. A precise computation of the numerical  
coefficients is dealt with in Section 4.

The structure of the linearized metric (2.3) is that of a sum of retarded  
multipolar waves, consisting of $p$ spatial derivatives (say) acting on  
monopolar waves $r^{-1} X(t-r)$,

$$ h_1 \approx \sum \partial_P [ r^{-1} X (t-r) ] \ . \eqno(3.1) $$
Our notation $\approx$ refers to the structure of the expression. By expanding
the derivatives $\partial_P$ (which act both on the pre-factor $r^{-1}$  
and on the retardation $t-r$) we get

$$ h_1 \approx \sum_{j\geq 1} \hat n_Q r^{-j} Z (t-r) \ , \eqno(3.2) $$

\noindent where the powers of $1/r$ are $j \geq 1$, and where we have  
expressed the angular dependence of each term using STF products of
unit vectors $\hat{n}_Q=n_{<i_{i}}n_{i_2}...n_{i_q>}={\rm STF}$ part of
$n_Q$ (see [31]). In practice, one may compute (3.2) from (3.1) by
decomposing $\partial_P$ on the basis of STF spatial derivatives  
$\hat{\partial}_Q$ and using (A.15) in Appendix A. After insertion of (3.2)
into the quadratic source term $N_2^{\mu\nu}$ one finds

$$ N_2 \approx \sum_{k\geq 2} \hat n_L r^{-k} F (t-r)\ , \eqno(3.3) $$

\noindent where the powers of $1/r$ start with $k=2$ (as is clear from
the fact that $N_2$ is quadratic in $h_1$ which is of order $1/r$). The
functions $F$ are composed of sums of quadratic products of derivatives
of the functions $Z$ in (3.2).

The main problem is to compute $u_2$ defined by (2.8). In view of the
structure (3.3), it is {\it a priori} required to compute the finite
part ${\rm FP}_{B=0}$ of the retarded integral of any term
$\hat{n}_Lr^{-k} F(t-r)$ with multipolarity $\ell$ and radial dependence  
with $k \geq 2$. All the required formulas are listed in Appendix A  
(which summarizes results obtained mostly in previous works [20,8,9]). 
Notably, we know from Appendix A that the (finite part of the) retarded  
integral in the case $k =2$ is irreducibly {\it non-local} or
non-instantaneous. See (A.3)--(A.7) in Appendix A. When the power $k$  
satisfies $3 \leq k \leq \ell+2$, where $\ell$ is the multipolarity  
(this excludes the monopolar case $\ell=0$), we know that the corresponding  
retarded integral is instantaneous, and given by (A.11)--(A.12). Finally,
when $k \geq \ell+3$, we have again a non-local expression, given  
by (A.13)--(A.14), except in special combinations like (A.16) for which  
the non-local integrals cancel out.

The computation of $u_2= {\rm FP}\square ^{-1}_R N_2$ can be implemented
by an algebraic computer program, following the successive steps
(3.1)--(3.3) and applying on each terms composing (3.3) the formulas
(A.5), (A.11) and (A.13)--(A.14). This is probably the most efficient way  
to obtain $u_2$. However, in doing so we would discover that the only
non-local integrals left in $u_2$ come from the source terms having a
radial dependence with $k=2$, in other words all the non-local integrals coming
from terms having $k \geq \ell+3$ actually cancel out. Practically speaking  
the source terms having $k \geq \ell+3$ turn out to combine into combinations
such as (A.16) yielding purely instantaneous contributions. This fact (proved
in [8]) is special to the quadratic non-linearities, and does not stay
true at higher orders, e.g. at the cubic order as seen in paper II.

The proof of the latter assertion uses specifically the quadratic
structure of the source $N_2$. Instead of inserting in (2.5) the
linearized metric in expanded form (3.2) and then working out all
derivatives to arrive at (3.3), we keep the structure of the source as
it basically is, i.e. a sum of quadratic products of multipolar waves,

$$ N_2 \approx \sum \partial_P \Bigl[ r^{-1} X (t-r) \Bigr] \partial_Q
   \Bigl[ r^{-1} Y(t-r) \Bigr] \ , \eqno(3.4) $$

\noindent involving spatial multi-derivatives with $P=i_1...i_p$ and
$Q=j_1...j_q$ and some functions of time $X$ and $Y$. Then we perform
on each term of (3.4) a sequence of operations by parts [i.e. $\partial_i A  
\partial_j B =\partial_i(A\partial_jB) -A\partial_i\partial_jB$], by which  
the spatial derivatives acting on the wave in the left (say) are shifted  
in front and to the right. This leads to

$$ N_2 \approx \sum \partial_P \Bigl\{ r^{-1} X (t-r) \partial_R
   [r^{-1} Y(t-r)] \Bigr\} \ . \eqno(3.5) $$
Only at this stage does one expand the space derivatives $\partial_R$
(inside the curly brackets), while leaving the derivatives $\partial_P$
in front un-expanded. The result reads

$$ N_2 \approx \sum_{2\leq k \leq \ell+2} \partial_P \left\{ \hat n_L r^{-k}
   H(t-r) \right\}  \ , \eqno(3.6) $$

\noindent where the functions $H$ are sums of products of $X$ and
time-derivatives of
$Y$, and where we have projected the angular
dependence of the terms inside the brackets on STF tensors $\hat{n}_L$. 
The point is that the radial dependence of the terms inside the brackets  
is related to the multipolarity $\ell$ by $2\leq k \leq \ell+2$. This can  
easily be seen from (3.5), as the expansion of the multipolar wave  
$\partial_R[r^{-1}Y]$ is composed of a sum of terms $\hat\partial_L
[r^{-1}Y']$ which have $1 \leq k \leq \ell+1$ [see (A.1)], yielding $2
\leq k \leq \ell+2$ after multiplication by the factor $r^{-1}$ on the
left.  The next operation is to single out the terms with pure radial
dependence $k=2$. All these terms can be obtained by applying $\partial_P$
on the terms inside the brackets having $k=2$.  In this way one
generates besides all the terms $k=2$ many other terms with $k \geq 3$,
but the latter terms can be recombined into terms of the same form as in
(3.6) (and thus having $3 \leq k \leq \ell+2$).  See [8] for the proof.
Thus (3.6) can be rewritten as

$$ N_2 \approx r^{-2} Q ({\bf n},t-r) + \sum_{3\leq k \leq \ell+2}
  \partial_P \left\{ \hat n_L r^{-k} F(t-r) \right\}\ , \eqno(3.7) $$

\noindent where $Q ({\bf n},t-r)$ denotes the coefficient of $r^{-2}$ in
the (finite) expansion of the quadratic source when $r \to \infty$ with
$t-r={\rm const}$. The definition of $Q ({\bf n},t-r)$ is

$$ N_2^{\mu\nu} = {1 \over r^2} Q^{\mu\nu} ({\bf n},t-r)
+ O\left( {1\over r^3} \right) \ . \eqno(3.8) $$
Next, in anticipation of applying the finite part of the retarded
integral, we multiply (3.7) by a factor $r^B$, and introduce $r^B$ inside  
the brackets using again a series of operations by parts. In this way
we get many new terms, but which all involve at least one factor $B$ coming  
from the differentiation of $r^B$ during the latter operations. Thus

$$ r^B N_2 \approx r^{B-2} Q ({\bf n},t-r) + \sum_{3\leq k \leq \ell+2}
  \partial_P \left\{ \hat n_L r^{B-k} F(t-r) \right\} + O(B) \ . \eqno(3.9)
$$
Applying the retarded integral on both sides of (3.9), commuting
$\square_R ^{-1}$ with $\partial_P$, and taking the finite part,
we are left with (the finite part of) retarded integrals of three types  
of terms: (i) the first term in (3.9) which has radial dependence $r^{-2}$, 
(ii) the terms in the brackets of (3.9) having radial dependence such  
that $3 \leq k \leq \ell+2$, and (iii) the terms $O(B)$ which have  
the structure (3.3) with any radial dependence $k$ but carry at least  
one factor $B$. In Case (ii), the retarded integrals are given by the  
instantaneous expressions (A.11). In Case (iii), the retarded integrals
are given by (A.18) when the power of $B$ is one, and are zero for higher  
powers. So in Case (iii) the retarded integrals are also instantaneous.
Therefore we can state the following result [8]: the only non-local integrals  
in $u_2={\rm FP}\square_R^{-1}N_2$ come from Case (i), namely from the
source terms whose radial dependence is $r^{-2}$, and which are denoted  
by $r^{-2} Q ({\bf n},t-r)$ in (3.8). These non-local integrals are
given by (A.3)--(A.7). This result holds true only in the case of  
the quadratic non-linearity. In cubic and higher non-linearities, some  
hereditary integrals are generated by source terms with radial dependence  
such that $k \geq \ell+3$.

Incidentally, note that the decomposition (3.9) of the source shows that the
$B$-dependent retarded integral $\square_R ^{-1}[r^B N_2]$ is {\it finite} 
at $B=0$, i.e. does not involve any pole when $B \to 0$. Thus the finite  
part at $B=0$ is simply equal to the value of $\square_R ^{-1}[r^BN_2]$
at $B=0$. This can be checked from the formulas (A.3), (A.11) and (A.18)  
in Appendix A, which are all finite at $B=0$. Here again this is a peculiarity of the quadratic approximation.

We are now in the position to write down the structure of the first
contribution $u_2$. From (3.9) and (A.3), (A.11) and (A.18) we have

$$ u_2 \approx t_2 + \sum_{k\geq 1} \hat n_L r^{-k} G(t-r) \ ,\eqno(3.10) $$

\noindent where the functions $G(t-r)$ depend instantaneously on the
multipole moments $M_L$ and $S_L$ (i.e. at the retarded time $t-r$
only), and where the first term is non-local and given by

$$ t^{\mu\nu}_2 = \square^{-1}_R [ r^{-2} Q^{\mu\nu} ({\bf n}, t-r)]\ .
         \eqno(3.11) $$
On the other hand the second contribution $v_2$, defined by (2.11)-(2.12), 
involves some time anti-derivatives of quadratic products of moments. 
We denote these time anti-derivatives by $s_2$. Thus the quadratic-order  
metric can be written as

$$ h_2 \approx t_2 + s_2 + \sum_{k\geq 1} \hat n_L r^{-k} P (t-r)
\ ,   \eqno(3.12) $$

\noindent where $t_2$ is the non-local integral (3.11), where $s_2$ are some  
anti-derivatives [given by (4.12) below in the case of the
quadrupole-quadrupole interaction], and where we have many instantaneous terms.
See (4.13) and Table~2 below for the complete expression of $h_2$ in the
case of the quadrupole-quadrupole interaction.

\section{The quadrupole-quadrupole metric}   

We specialize the previous investigation in the case of the interaction
between two quadrupole moments $M_{ab}$ and $M_{cd}$. Thus we keep in
the linearized metric (2.3) only the terms corresponding to $M_{ab}$,

$$ \eqalignno{
  h^{00}_1 &= -2 \partial_{ab} \left( r^{-1} M_{ab} \right) \ , &(4.1a)\cr
  h^{0i}_1 &= 2 \partial_{a} \left( r^{-1} M^{(1)}_{ai} \right) \ , &(4.1b)\cr
  h^{ij}_1 &= -2 r^{-1} M^{(2)}_{ij} \ . &(4.1c)\cr } $$
(Henceforth we use the same notation for the metric constructed out of
the quadrupole moment as for the complete metric involving all multipolar  
contributions.)

Inserting (4.1) into the quadratic source (2.5), we can work out
explicitly all the terms composing the source either in the all-expanded
form (3.3) or in the more elaborate form (3.7). Notably, we find that
the terms with radial dependence $r^{-2}$ take the classic form of the
stress-energy tensor of a massless field,

$$ Q^{\mu\nu} ({\bf n},t-r)= k^\mu k^\nu \Pi ({\bf n},t-r)\ , \eqno(4.2) $$

\noindent where $k^\mu$ denotes the Minkowskian null vector $k^\mu=(1,n^i)$,
and where $\Pi$ is given by

$$ \Pi =n_{abcd} M^{(3)}_{ab} M^{(3)}_{cd} - 4n_{ab} M^{(3)}_{ac} M^{(3)}_{bc}
       + 2 M^{(3)}_{ab} M^{(3)}_{ab} \ .   \eqno(4.3) $$
The quantity $\Pi$ is proportional to the power (per unit of steradian)  
carried by the linearized waves. In the general case, $Q^{\mu\nu}$  
involves besides the quadrupole-quadrupole terms all the interacting
terms between multipole moments with $\ell \geq 2$ and the mass
monopole $M$. We introduce the STF multipole decomposition of $\Pi$,

$$ \Pi ({\bf n}, u) = \sum_{\ell \geq 0} n_L\, \Pi_L (u) \ . \eqno(4.4) $$

\noindent From (4.3), the only non-zero multipolar coefficients are

$$ \eqalignno{
      \Pi_0 &= {4\over 5}\, M^{(3)}_{ab} M^{(3)}_{ab} \ , &(4.5a) \cr
   \Pi_{ij} &= -{24\over 7}\, M^{(3)}_{a<i} M^{(3)}_{j>a} \ , &(4.5b) \cr
   \Pi_{ijkl} &=  M^{(3)}_{<ij} M^{(3)}_{kl>} \ . &(4.5c) \cr } $$

\noindent ($\Pi_0$ denotes the coefficient with multipolarity $\ell=0$
or spherical average.) With this definition, and with the help of (A.7)
in Appendix A, we can write the non-local integral $t_2$ as

$$ t^{\mu\nu}_2 = \square^{-1}_R \left[ {k^\mu k^\nu\over r^2}\Pi  \right]
    = - \int^{+\infty}_r ds \int {d\Omega'\over 4\pi}
{k'^\mu k'^\nu\over s-r{\bf n}.{\bf n}'} \Pi ({\bf n}', t-s)\ .\eqno(4.6) $$
Some equivalent expressions follow from (A.3)--(A.5). For instance,
we can write $t^{\mu\nu}_2$ in the explicit form

$$t^{\mu\nu}_2 = {1\over 2} \int ^{+\infty}_r ds \left[ M^{(3)}_{ij}
  M^{(3)}_{kl} \partial^{\mu\nu}_{ijkl} \{6\} - 4M^{(3)}_{ai} M^{(3)}_{aj}
  \partial^{\mu \nu}_{ij} \{4\} + 2 M^{(3)}_{ab} M^{(3)}_{ab}
    \partial^{\mu \nu} \{2\} \right]  \ , \eqno(4.7a)$$
where the moments in the integrand are evaluated at the time $t-s$, where
the multi-derivative operators mean for instance $\partial_{ij}^{\mu
\nu}=\partial^\mu \partial^\nu \partial_i \partial_j$ with
$\partial^\mu =(-\partial/\partial s, \partial_i)$, and where we use the
special notation

$$\{p\}={(s-r)^p \ln (s-r) - (s+r)^p \ln (s+r) \over p!r}   \eqno(4.7b)$$
[see (A.4) in Appendix A].

Having the term $t_2^{\mu \nu}$, we undertake the computation of all the  
instantaneous terms $M_{ab} \times M_{cd}$ in $u_2^{\mu \nu}$
[second terms in (3.10)]. The computation is straigthforward but
tedious. As said above, when doing practical computations (notably by
computer), the best method is the somewhat brute force method consisting of
obtaining the source in the all-expanded form (3.3), and applying the
(finite part of the) retarded integral on each term of (3.3) with $k\geq3$
using the formulas (A.11) and (A.13). This is simpler than working out the
source in the more elaborate form (3.9), and using the manifestly
instantaneous formulas (A.11) and (A.18). The brute force method has also
the advantage that one can {\it check} that all the non-local integrals
but those coming from the $r^{-2}$ term cancel out (as well as the  
associated logarithms of $r$). The term $u_2$ takes the form

$$  \eqalignno{
u^{00}_2 &= t^{00}_2 + \sum^6_{k=1} {1\over r^k} \sum^{6-k}_{m=0}
\biggl\{a^k_m \hat n_{abcd}M^{(6-k-m)}_{ab}M^{(m)}_{cd} \cr
  &\qquad\qquad\qquad\qquad + b^k_m \hat n_{ab}M^{(6-k-m)}_{ac} M^{(m)}_{bc}
   + c^k_mM^{(6-k-m)}_{ab}M^{(m)}_{ab} \biggr\}    \ , &(4.8a)  \cr
u^{0i}_2 &= t^{0i}_2 + \sum^6_{k=1} {1\over r^k} \sum^{6-k}_{m=0}
   \biggl\{d^k_m \hat n_{iabcd} M^{(6-k-m)}_{ab}M^{(m)}_{cd}  \cr
  &\qquad\qquad\qquad\qquad + e^k_m \hat n_{iab} M^{(6-k-m)}_{ac}M^{(m)}_{bc}
   + f^k_m n_i M^{(6-k-m)}_{ab}M^{(m)}_{ab}  \cr
  &\qquad\qquad\qquad\qquad + g^k_m \hat n_{abc} M^{(6-k-m)}_{ia}M^{(m)}_{bc}
   + h^k_m n_a M^{(6-k-m)}_{ib} M^{(m)}_{ab} \biggr\} \ ,   &(4.8b) \cr
u^{ij}_2 &= t^{ij}_2 + \sum^6_{k=1} {1\over r^k} \sum^{6-k}_{m=0}
 \biggl\{p^k_m \hat n_{ijabcd}M^{(6-k-m)}_{ab}M^{(m)}_{cd}  \cr
  &\qquad\qquad\qquad\qquad + q^k_m \hat n_{ijab} M^{(6-k-m)}_{ac}M^{(m)}_{bc}
  + r^k_m \delta_{ij} \hat n_{abcd} M^{(6-k-m)}_{ab} M^{(m)}_{cd} \cr
  &\qquad\qquad\qquad\qquad + s^k_m \hat n_{ij} M^{(6-k-m)}_{ab}M^{(m)}_{ab}
   + t^k_m \delta_{ij} \hat n_{ab}M^{(6-k-m)}_{ac}M^{(m)}_{bc} \cr
  &\qquad\qquad\qquad\qquad + u^k_m \delta_{ij} M^{(6-k-m)}_{ab}M^{(m)}_{ab}
   + v^k_m \hat n_{abc(i}M^{(6-k-m)}_{j)a} M^{(m)}_{bc} \cr
  &\qquad\qquad\qquad\qquad + w^k_m \hat n_{a(i}M^{(6-k-m)}_{j)b}M^{(m)}_{ab}
   + x^k_m n_{ab} M^{(6-k-m)}_{ij}M^{(m)}_{ab} \cr
  &\qquad\qquad\qquad\qquad + y^k_m \hat n_{ab}M^{(6-k-m)}_{a(i} M^{(m)}_{j)b}
   + z^k_m M^{(6-k-m)}_{a(i}M^{(m)}_{j)a} \biggr\} \ ,   &(4.8c) \cr }
$$

\noindent where all moments are evaluated at time $t-r$ and where  
$a_m^k$, $b_m^k$, ..., $z_m^k$ are purely numerical coefficients.
Using the algebraic computer program Mathematica [32] we have obtained
the numerical coefficients listed in Table 1.

Next we follow the second part of the construction of the metric, and
compute the divergence $w_2^\mu = \partial_\nu u_2^{\mu \nu}$ [see
(2.10)-(2.12)]. To this end we need the divergence of the integral  
$t_2^{\mu \nu}$, which is easily evaluated by noticing that  
$\partial_\nu t^{\mu \nu}_2$ can be written, analogously to (2.10), as some  
residue at $B=0$ of a retarded integral (because of the explicit factor  
$B$), and furthermore that the radial dependence of the integrand is merely  
$r^{-3}$ (because it comes from the differentiation of the source term  
$r^{-2}$). From (A.18), we know that when $k=3$ the residue is non-zero  
only when the multipolarity is $\ell=0$. This yields immediately

$$ \partial_\nu t^{\mu\nu}_2 = \hbox{FP}_{B=0}\, \square^{-1}_R
   \left[ B \left( {r\over r_0} \right)^{\!B} {k^\mu\over r^3} \Pi \right]
 = - {1\over r}\int {d\Omega\over 4\pi} k^\mu \Pi ({\bf n},t-r) \ . \eqno(4.9) $$
The $\mu=0$ component of (4.9) is proportional to the angular average of
$\Pi$, already computed in (4.5a). One can check that the $\mu=i$
component is zero. Thus,

$$ \eqalignno{
  \partial_\nu t^{0\nu}_2 &= -{4\over 5} r^{-1} M^{(3)}_{ab} M^{(3)}_{ab}
  \ , & (4.10a) \cr
  \partial_\nu t^{i\nu}_2 &= 0 \ . & (4.10b) \cr } $$
Knowing (4.10) we can obtain $w_2^\mu$ by direct differentiation of the
expressions (4.8), using the coefficients in Table 1. Again the
computation is quite lengthy, but it provides us with an important check.
Indeed, from (2.11) one must find that the divergence $w_2^\mu$ is
a solution of the source-free d'Alembertian equation, namely  
$\square w_2^\mu=0$. This test is very stringent, as a single erroneous  
coefficient in Table 1 would signify almost certainly its failure. 
Thus we determine all the tensors $A_L$, ..., $D_L$ in (2.11). 
The non-zero ones are given by

$$ \eqalignno{
A &=-{2\over 75} M^{(6)}_{ab} M_{ab} -{176\over 225} M^{(5)}_{ab} M^{(1)}_{ab}
 -{22\over 45} M^{(4)}_{ab} M^{(2)}_{ab} -{8\over 15} M^{(3)}_{ab} M^{(3)}_{ab}
  \ , &(4.11a) \cr
A_{ij} &= -{24\over 35} \left( M^{(4)}_{a<i} M_{j>a}
     +M^{(3)}_{a<i} M^{(1)}_{j>a} \right) \ , &(4.11b) \cr
B &= -{32\over 75} M^{(5)}_{ab} M_{ab} -{8\over 5} M^{(4)}_{ab} M^{(1)}_{ab}
    +{8\over 15} M^{(3)}_{ab} M^{(2)}_{ab}  \ , &(4.11c) \cr
C_{ij} &= {4\over 25} M^{(5)}_{a<i} M_{j>a}
    -{4\over 5} M^{(4)}_{a<i} M^{(1)}_{j>a}
    +{8\over 5} M^{(3)}_{a<i} M^{(2)}_{j>a}   \ , & (4.11d) \cr
D_i &= \varepsilon_{iab} \left( -{2\over 5} M^{(5)}_{ac} M_{bc}
   -2 M^{(4)}_{ac} M^{(1)}_{bc} - {4\over 5} M^{(3)}_{ac} M^{(2)}_{bc}\right)
\ .  &(4.11e)\cr } $$
By inserting the latter values into (2.12), we obtain the components of the  
second part $v_2^{\mu \nu}$. In particular, we find some time anti-derivatives  
which define $s_2^{\mu \nu}$ [see (3.12)] as

$$ \eqalignno{
  s^{00}_2 &= {4\over 5} r^{-1} \int^{t-r}_{-\infty} du M^{(3)}_{ab}
     M^{(3)}_{ab} (u) \ ,  &(4.12a) \cr
  s^{0i}_2 &= -{4\over 5} \varepsilon_{iab} \partial_a  \left(
   r^{-1} \varepsilon_{bcd} \int^{t-r}_{-\infty} du M^{(3)}_{ce}
     M^{(2)}_{de} (u) \right) \ , &(4.12b) \cr
  s^{ij}_2 &= 0  \ . &(4.12c) \cr }
$$
The physical interpretation of these anti-derivatives is clear. Indeed the
linearized metric (2.3) depends in particular on the mass monopole $M$
and current dipole $S_i$ of the source (both are constant). Physically
$M$ and $S_i$ represent the total mass-energy and angular momentum of
the system before the emission of gravitational radiation. Now the
integral $s^{00}_2$ given by (4.12a) represents a small modification, due  
to the emission of radiation, of the initial mass $M$. This is clear from  
the comparison of (4.12a) and (2.3a), showing that there is exact agreement  
with the energy loss by radiation as given by the standard Einstein  
quadrupole formula. This result is originally due to Bonnor [10], 
and Bonnor and Rotenberg [4]. Similarly the integral $s^{0i}_2$
given by (4.12b) represents a modification of the total angular momentum  
in agreement with the quadrupole formula for the angular momentum loss.
[There is no loss of total linear momentum at the level of the
quadrupole-quadrupole interaction (one needs to consider also the mass
octupole $M_{abc}$ and/or the current quadrupole $S_{ab}$).]

The quadratic metric $h_2^{\mu \nu}$ can now be completed. 
We add up the two contributions $u_2^{\mu \nu}$ [given by (4.8) and Table 1]  
and $v_2^{\mu \nu}$ [given by (4.11) and (2.12)]. The local terms are written
in the same form as in (4.8). Thus,

$$ \eqalignno{
   h^{\mu\nu}_2 = t^{\mu\nu}_2 + s^{\mu\nu}_2 + \sum^6_{k=1}{1\over r^k}
    \sum^{6-k}_{m=0} \biggl\{ &
    \hbox{same expressions as in (4.8abc) but} \cr
    &\hbox{with coefficients}\ a'^k_m, ..., z'^k_m \biggr\}
   \ , &(4.13) \cr } $$

\noindent where the non-local integrals $t_2^{\mu \nu}$ and $s_2^{\mu\nu}$
are given by (4.7) and (4.12), and where all the coefficients
$a'^k_m, ..., z'^k_m$ are listed in Table 2.

\section{The quadrupole-quadrupole metric in the far zone} 

We investigate the behaviour of the quadrupole-quadrupole field
$h_2^{\mu\nu}$ in the far zone, near future null infinity (i.e.  at
large distances when we recede from the source at the speed of light).
The degrees of freedom of the radiation field, at leading order in the inverse
of the distance, are contained in the transverse and tracefree (TT) projection  
of the spatial components $ij$ of the metric (say $g^{\rm TT}_{ij}$).
These are the so-called observable (or radiative) multipole moments,
which are ``measured'' in an experiment located far away from the
system. The TT projection $g^{\rm TT}_{ij}$ to first order in $1/r$
reads 

$$ \eqalignno{
 g^{\rm TT}_{ij} = {4 \over r} {\cal P}_{ijab}
 \sum^\infty_{\ell =2} {1\over \ell !} \biggl\{ n_{L-2} U_{ijL-2}
 - {2\ell \over {\ell +1}} n_{aL-2}
 \varepsilon_{ab(i} V_{j)bL-2} \biggr\} +
 O \left( {1\over r^2}\right)  \cr
    & & (5.1) \cr } $$

\noindent (with $G=c=1$), where $U_L$ and $V_L$ denote the mass-type and
current-type observable moments (both are functions of $t-r$), and where
the TT projection operator is

$$ {\cal P}_{ijab} ({\bf n}) = (\delta_{ia} -n_in_a) (\delta_{jb} -n_jn_b)
-{1\over 2} (\delta_{ij} -n_in_j) (\delta_{ab} -n_an_b) \ . \eqno(5.2) $$
The $\ell$-dependent coefficients in (5.1) are chosen so that $U_L$ and
$V_L$ agree at the linearized order with the $\ell$th time derivatives
of the moments $M_L$ and $S_L$ [compare with (2.3c)].

Let us consider first the non-local integral $t_2^{\mu \nu}$
[see (4.6)-(4.7)]. As we know from (A.8), the asymptotic expansion when  
$r\to \infty$, $t-r={\rm const}$ of the retarded integral of a source  
with radial dependence $r^{-2}$ is composed of terms $1/r^n$ and $\ln r/r^n$. 
As such, $t_2^{\mu \nu}$ behaves like $\ln r/r$ when $r \to\infty$,
$t-r=$~const. The logarithm is due to the deviation of the flat cones
$t-r=$~const in harmonic coordinates from the true space-time null cones.
The metric in harmonic coordinates is not of the normal Bondi-type at
future null infinity [33,34]. Removal of the logarithm is done using
radiative coordinates, so defined that the associated flat cones agree,  
asymptotically when $r\rightarrow +\infty$, with the true null cones
(see e.g.  [21]). This method, adopted in [9], permits to compute the
observable moments in the non-local term $t_2^{\mu \nu}$. Here we follow  
another method, found by Thorne [15] and Wiseman and Will [16], which
consists of applying first the TT projection operator (5.2) on $t_2^{\mu \nu}$. 
Because the TT projection kills any (linear) gauge term in the $1/r$ part
of the metric, this method shortcuts the need of a transformation
to radiative coordinates. However, one must be cautious in taking the
limit $r\to \infty$, $t-r=$~const using (4.6).
It is not allowed, for instance, to work out a leading $1/r$ term
from the second expression in (4.6) because this term would involve
a divergent integral (in accordance with the fact that the leading term  
is actually $\ln r/r$). But, as pointed out in [15,16], the divergent parts of  
the integral cancel out after application of the TT projection, and at the end  
one recovers the correct result.
Thus, we compute

$$ (t^{ij}_2)^{\rm TT} = - {\cal P}_{ijab} ({\bf n}) \int^{+\infty}_r ds \int
  {d\Omega'\over 4\pi} {n'^a n'^b\over s-r{\bf n}.{\bf n}'}\, \Pi
  ({\bf n}', t-s) \ . \eqno(5.3) $$

\noindent [Note that the TT projection as defined in (5.2) is purely  
algebraic. Strictly speaking it agrees with the true TT projection  
only when acting on the leading $1/r$ term.] With the multipole  
decomposition (4.4) we have

$$ (t^{ij}_2)^{\rm TT} = - {\cal P}_{ijab} ({\bf n}) \sum_{\ell \geq 0}
\int^{t-r}_{-\infty} du \, \Pi_L (u) \int {d\Omega'\over 4\pi} {n'_a
n'_b n'_L\over t-u-r{\bf n}.{\bf n}'} \ .  \eqno(5.4) $$
We decompose the product of unit vectors $n'_a n'_b n'_L$ on the
basis of STF tensors [31], we drop the terms having zero TT-projection,
and we express the remaining terms using the Legendre function of the  
second kind $Q_\ell$ [see (A.6c)]. Restoring the traces on the STF tensors, 
and droping further terms having zero TT-projection, we obtain

$$ \eqalignno{
 (t^{ij}_2)^{\rm TT} &= -{1\over r} {\cal P}_{ijab} \sum_{\ell\geq 2}
  {\ell (\ell-1) \over (2\ell+3)(2\ell+1)(2\ell-1)} n_{L-2}
  \int^{t-r}_{-\infty} du\, \Pi_{abL-2} (u)  \cr
  &\times \biggl\{ (2\ell-1) Q_{\ell+2} \left( {t-u\over r} \right)
  - 2(2\ell+1) Q_\ell \left( {t-u\over r} \right)
  + (2\ell+3) Q_{\ell-2} \left( {t-u\over r} \right) \biggr\}   \ . \cr
  & &(5.5) \cr } $$
The limit at future null infinity can now be applied. Indeed it suffices to
insert the expansion of $Q_\ell(x)$ when $x\to 1^+$ as given by (A.8).
As expected the limit is finite, because the terms $\ln(x-1)/2$ in the  
expansions of the $Q_\ell$'s cancel out. This yields [13,15,16,9]

$$ (t^{ij}_2)^{\rm TT} = -{2\over r} {\cal P}_{ijab} \sum_{\ell\geq 2}
   {n_{L-2}\over (\ell+1)(\ell+2)} \int^{t-r}_{-\infty} du \,
   \Pi_{abL-2} (u) + O \left( {\ln r\over r^2} \right) \ . \eqno(5.6) $$
In the case of the quadrupole-quadrupole interaction we find [using (4.5)]

$$ (t^{ij}_2)^{\rm TT} = {1\over r} {\cal P}_{ijab} \int^{t-r}_{-\infty} du
   \left[ {4\over 7} M^{(3)}_{c<a} M^{(3)}_{b>c} - {1\over 15} n_{cd}
   M^{(3)}_{<ab} M^{(3)}_{cd>} \right]
   + O \left({\ln r\over r^2}\right) \eqno(5.7) $$

\noindent (where the brackets $<>$ denote the STF projection). The
non-local integral (5.7) represents the quadrupole-quadrupole
contribution to the non-linear memory [14-16,9].

We now turn our attention to the instantaneous part of the metric. 
All the terms have been obtained in Section 4. We need only to apply the TT
projection on the $1/r$ part of $h_2^{ij}$ as given by (4.13) and the  
coefficients listed in Table 2. After several transformations using STF 
techniques we obtain

$$ \eqalignno{
  (h^{ij}_2 - t^{ij}_2)^{\rm TT} &= {1\over r} {\cal P}_{ijab} \biggl\{
   - {2\over 7} M^{(5)}_{c<a} M_{b>c}
   + {10\over 7} M^{(4)}_{c<a} M^{(1)}_{b>c}
   + {4\over 7} M^{(3)}_{c<a} M^{(2)}_{b>c}  \cr
 &\quad\qquad\qquad + n_{cd} \biggl[ {7\over 10} M^{(5)}_{<ab} M_{cd>}
   + {21\over 10} M^{(4)}_{<ab} M^{(1)}_{cd>}
   + {17\over 5} M^{(3)}_{<ab} M^{(2)}_{cd>} \biggr] \cr
 &\quad\qquad\qquad + n_{dg} \varepsilon_{acg} \left(
   \varepsilon_{ef<b} \left[ {1\over 10} M^{(5)}_{c {\underline e}} M_{d>f}
 - {1\over 2} M^{(4)}_{c{\underline e}} M^{(1)}_{d>f}\right]\right)\biggr\}\cr
  &\qquad + O\left( {1\over r^2} \right) & (5.8) \cr }
$$

\noindent (where the overbar on $\underline e$ means that the index $e$
is to be excluded from the STF projection).

With (5.7) and (5.8) we deduce the observable moments by comparison with  
(5.1). The quadrupole-quadrupole interaction contributes only to the  
observable mass quadrupole moment $U_{ij}$, mass $2^4$-pole moment  
$U_{ijkl}$, and current octupole moment $V_{ijk}$. We find

$$ \eqalignno{
 \delta U_{ij} &= -{2\over 7} \int^{t-r}_{-\infty} \,
    M^{(3)}_{a<i} M^{(3)}_{j>a}
   + {1\over 7} M^{(5)}_{a<i} M_{j>a}
  -{5\over 7} M^{(4)}_{a<i} M^{(1)}_{j>a}
   - {2\over 7} M^{(3)}_{a<i} M^{(2)}_{j>a} \ ,\cr
 & &(5.9a)  \cr
 \delta U_{ijkl} &= {2\over 5} \int^{t-r}_{-\infty} \,
    M^{(3)}_{<ij} M^{(3)}_{kl>}
   - {21\over 5} M^{(5)}_{<ij} M_{kl>}
   -{63\over 5} M^{(4)}_{<ij} M^{(1)}_{kl>}
   - {102\over 5} M^{(3)}_{<ij} M^{(2)}_{kl>}\ ,  \cr
  & &(5.9b)  \cr
 \delta V_{ijk} &=
   \varepsilon_{ab<i} \left[ {1\over 10} M^{(5)}_{j{\underline a}} M_{k>b}
   - {1\over 2} M^{(4)}_{j{\underline a}} M^{(1)}_{k>b} \right]
 \ . & (5.9c) \cr } $$
Note that the non-local integrals are present in $U_{ij}$ and $U_{ijkl}$, 
but not in the current moment $V_{ijk}$.

Finally we add back in (5.9) the factor $G$ and the powers of $1/c$ which are  
required in order to have the correct dimensionality. When this is
done we find that $\delta U_{ij}$ is of order $1/c^5$ in the post-Newtonian  
expansion (2.5PN order), while both $\delta U_{ijkl}$ and $\delta V_{ijk}$  
are of order $1/c^3$ or 1.5PN. This permits writing the complete  
expressions of $U_{ij}$ to 2.5PN order, and $U_{ijkl}$, $V_{ijk}$ to 1.5PN
order. In the case of $U_{ij}$ the reasoning has been done in [25],
which shows that besides the quadrupole-quadrupole terms $M_{ab}\times M_{cd}$
there is an interaction between $M_{ab}$ and the (static) current dipole  
$S_c$ also at 2.5PN order, and there is the standard contribution of tails  
(computed in [9]) at 1.5PN order. [In principle there is also an interaction  
between the mass octupole $M_{abc}$ and the mass dipole $M_d$ at 2.5PN order, 
but we have chosen $M_d=0$.] Thus, from (5.9a) and the equation (5.7) in [25], 

$$  \eqalignno{
  U_{ij} (t-r/c) =&~M^{(2)}_{ij} + {2GM\over c^3} \int^{t-r/c}_{-\infty}
   du \left[ \ln \left( {t-r/c -u\over 2b} \right) + {11\over 12} \right]
   M^{(4)}_{ij} (u) \cr
  & +{G\over c^5} \biggl\{ -{2\over 7} \int^{t-r/c}_{-\infty} du
     M^{(3)}_{a<i} M^{(3)}_{j>a} (u)
   + {1\over 7} M^{(5)}_{a<i} M_{j>a}   \cr
  & -{5\over 7} M^{(4)}_{a<i} M^{(1)}_{j>a}
   - {2\over 7} M^{(3)}_{a<i} M^{(2)}_{j>a}
   + {1\over 3} \varepsilon_{ab<i} M^{(4)}_{j>a} S_b \biggr\}
   +O \left( {1\over c^6} \right) \ . \cr
  &    & (5.10) \cr }
$$
The tail integral involves a constant 11/12 computed in Appendix B of
[9]. (See also [9] for the definition of the constant $b$.)
The coefficient in front of the term $M_{ab}\times S_c$ is computed  
in Appendix B below. In a similar way, we find that $U_{ijkl}$ and $V_{ijk}$  
involve the same types of multipole interactions, but that the terms  
$M_{ab}\times M_{cd}$ are of the same 1.5PN order as the tail terms. Thus,

$$ \eqalignno{
 U_{ijkl} (t-r/c) = M^{(4)}_{ijkl} + {G\over c^3} &\biggl\{ 2M
   \int^{t-r/c}_{-\infty} du
   \left[ \ln \left( {t-r/c -u\over 2b} \right) + {59\over 30} \right]
   M^{(6)}_{ijkl} (u) \cr
  & +{2\over 5} \int^{t-r/c}_{-\infty} du M^{(3)}_{<ij} M^{(3)}_{kl>} (u)
    - {21\over 5} M^{(5)}_{<ij} M_{kl>}  \cr
  & -{63\over 5} M^{(4)}_{<ij} M^{(1)}_{kl>}
    - {102\over 5} M^{(3)}_{<ij} M^{(2)}_{kl>} \biggr\}
 +O \left( {1\over c^4} \right)\ , \cr
  &  &(5.11) \cr
  V_{ijk} (t-r/c) = S^{(3)}_{ijk} + {G\over c^3} &\biggl\{ 2M
   \int^{t-r/c}_{-\infty} du
   \left[ \ln \left( {t-r/c -u\over 2b} \right) + {5\over 3} \right]
   S^{(5)}_{ijk} (u) \cr
  & +\varepsilon_{ab<i} \left[ {1\over 10} M^{(5)}_{j \underline{a}} M_{k>b}
    - {1\over 2} M^{(4)}_{j \underline{a}} M^{(1)}_{k>b}  \right]   \cr
  & - 2 S_{<i} M^{(4)}_{jk>} \biggr\} + O \left( {1\over c^4} \right)\ .
       &   (5.12) \cr  }
$$
The constants $59/30$ and $4/3$ are obtained from Appendix C in [24].
The coefficient of the term $S_{<i} M^{(4)}_{jk>}$ is computed in
Appendix B below.

The obtention in (5.10) of the observable quadrupole moment $U_{ij}$ up
to the 2.5PN order yields the total power contained in the radiation  
field (or energy flux) complete up to the same 2.5PN order. Indeed it
suffices to insert into (5.10) the intermediate quadrupole moment $M_{ij}$
determined in [25] as an explicit integral extending over the source at
the 2.5PN order (the other needed moments being the octupole $M_{ijk}$  
and current quadrupole $S_{ij}$ at 1PN order, and $M_{ijkl}$ and $S_{ijk}$  
at Newtonian order, which are all known). Note that in order to obtain
the wave form itself (and not only the energy flux it contains) one needs  
$M_{ijk}$ and $S_{ij}$ at the higher 2PN order.

\appendix{Formulas to compute the quadratic non-linearities} 

This appendix presents a unified compendium of formulas, many of them
issued from previous works [20,8,9], which permit the computation
of the quadratic non-linearities (involving any interaction between  
two multipole moments). The source of the quadratic non-linearities  
takes the structure (3.3), so we present the formulas for the finite  
part [as defined in (2.8)] of the retarded integral of any term  
$\hat{n} _L r^{-k} F(t-r)$ with multipolarity $\ell$ and radial  
dependence $r^{-k}$ where $k$ is an integer $\geq 2$. [For practical
computations it is convenient to use the source in expanded form (3.3)  
rather than in the more precise form (3.9).] The problem was solved  
in [20] which obtained a basic formula for the B-dependent retarded integral

$$ \eqalignno{
 &\square^{-1}_R \biggl[ \left( {r\over r_0}\right)^B {\hat n_L \over r^k}
  F(t-r) \biggr] =
 {2^{k-3} \over (2r_0)^B(B-k+2)(B-k+1)\cdots (B-k-\ell+2)} \cr
  &\qquad\quad \times \int^{+\infty}_r ds\, F(t-s)
  \hat\partial_L \left\{ {(s-r)^{B-k+\ell+2}
    -(s+r)^{B-k+\ell+2} \over r}\right\}\ .
     &({\rm A}.1) \cr } $$
This formula is valid (by analytic continuation) for all values of $B$
in the complex plane, except possibly at integer values of $B$ where there  
is a simple pole. Note that to the STF product of unit vectors $\hat{n} _L$  
in the left side corresponds a STF product of spatial derivatives
$\hat{\partial}_L$ in the right side [31].

Our first case of interest is that of a source having $k=2$. In this case  
(A.1) becomes

$$   \eqalignno{
 \square^{-1}_R &\biggl[ \left( {r\over r_0}\right)^B {\hat n_L \over r^2}
 F(t-r) \biggr] =
 {1 \over 2(2r_0)^B B(B-1)\cdots (B-\ell)}  \cr
 &\qquad\qquad\times \int^{+\infty}_r ds\, F(t-s)
  \hat\partial_L \left\{ {(s-r)^{B+\ell}
    -(s+r)^{B+\ell} \over r}\right\} \ , &({\rm A}.2)  \cr }  $$

\noindent of which we compute the finite part in the Laurent expansion
when $B\to 0$. We repeat briefly the reasoning of [8]:  the coefficient in
front of the integral admits a simple pole at $B=0$, but at the same
time the integral vanishes at $B=0$ thanks to the identity (A36) in [20]
(see also (4.20a) in [8]). As a result the right side of (A.2) is
finite at $B=0$, in agreement with the fact that the retarded integral
in its usual form (2.9) is convergent, with value

$$  \eqalignno{
  \square^{-1}_R &\left[ {\hat n_L \over r^2} F(t-r) \right] =
   {(-)^\ell\over 2} \int^{+\infty}_r ds\, F(t-s) 
   \cr &\qquad\qquad\times\hat\partial_L
   \left\{ {(s-r)^{\ell} \ln(s-r)
   -(s+r)^{\ell} \ln (s+r) \over \ell !r}\right\}\ , &({\rm A}.3) \cr } $$
where we have removed the reference to taking the finite part at $B=0$. 
Note that the length scale $r_0$ drops out in the result (this is thanks  
to (4.20a) in [8]). The formula (A.3) can be generalized to the case where  
the angular dependence is contained in any (non-tracefree) product  
$k_{\alpha_1}...k_{\alpha_\ell}$ of $\ell$ Minkowskian null vectors
$k_\alpha=(-1,{\bf n})$,

$$  \eqalignno{
\square^{-1}_R &\left[ {k_{\alpha_1}...k_{\alpha_\ell} \over r^2} F(t-r)
     \right] = {(-)^\ell\over 2} \int^{+\infty}_r ds\, F(t-s)
   \cr
&\qquad\qquad\times\partial_{\alpha_1}...\ \partial_{\alpha_\ell}
   \left\{ {(s-r)^{\ell} \ln(s-r)
   -(s+r)^{\ell} \ln (s+r) \over \ell !r}\right\}\ , 
   &({\rm A}.4)\cr }  $$
where the space-time derivatives in the right side are defined by
$\partial_\alpha=(\partial/\partial s, \partial_i)$. A useful
alternative form of (A.3), proved e.g. in Appendix A of [9], reads

$$ {\hbox{\rlap{$\sqcap$}$\sqcup$}}^{-1}_R \left[{\hat n_L \over r^2}
  F (t- r) \right] = - {\hat n_L \over r} \int^{+\infty}_r ds F (t-s)
    Q_{\ell} \left({s \over r} \right)\ , \eqno({\rm A}.5)  $$

\noindent where $Q_\ell(x)$ denotes the $\ell$th-order Legendre function
of the second kind (with branch cut from $-\infty$ to $1$, so $x>1$).
The Legendre function is given by

$$ \eqalignno{
 Q_{\ell} (x) &= {1 \over 2} \int^1_{-1} P_{\ell} (y) {dy\over x-y}
         &({\rm A}.6a) \cr
   &= {1\over 2} P_\ell (x) {\rm ln}
    \left({x+1 \over x-1} \right)- \sum^{ \ell}_{ j=1}
    {1 \over j} P_{\ell -j}(x) P_{j-1}(x) \ , &({\rm A}.6b)\cr } $$

\noindent where $P_\ell$ is the Legendre polynomial (see e.g.  [35]).
Note that by combining (A.6a) and the expansion of $1/(x-{\bf n}.{\bf n'})$  
in terms of Legendre polynomials (see e.g. (A26) in [20]), one has

$$  \hat n_L Q_\ell (x) = \int {d\Omega'\over 4\pi}
     {\hat n'_L\over x - {\bf n}.{\bf n}'} \ , \eqno({\rm A}.6c)  $$
where the angular integration $d\Omega'$ is associated with the unit
direction $n'^i$. Combining (A.5) and (A.6c) we obtain another alternative
formula,

$$ \square^{-1}_R \left[ {1\over r^2} F ({\bf n}, t-r) \right] =
  - \int^{+\infty}_r ds \int {d\Omega'\over 4\pi}
    {F ({\bf n}',t-s) \over s-r{\bf n}.{\bf n}'} \ . \eqno({\rm A}.7) $$
This formula is valid for any function $F({\bf n}, u)$ [not only for a
function having a definite multipolarity $\ell$ like in (A.5)]. It can
also be recovered from the retarded integral in its usual form (2.9).

The leading terms in the expansion when $r\to \infty$ (with $t-r=$~const)  
follow from the expansion of the Legendre function when $x\to 1^+$.  
The expression (A.6b) yields

$$ Q_\ell (x)= -{1\over 2} \ln \left( {x-1\over 2}\right) -
    \sum^\ell_{j=1} {1\over j} + O [(x-1) \ln (x-1)]  \ , \eqno({\rm A}.8a) $$

\noindent and with (A.5) this implies [9]

$$ \eqalignno{
  {\hbox{\rlap{$\sqcap$}$\sqcup$}}^{-1}_R \left[ {\hat n_L \over r^2} F
   (t-r) \right]
  &= {\hat n_L\over 2r} \int^{+\infty}_0 d\lambda
  F (t-r-\lambda)
    \biggl[\ln \left( {\lambda \over 2r} \right)
   + \sum^{\ell}_{j=1}{2 \over j} \biggr] \cr
  &+ O \left({\ln r \over r^2} \right)\ . &({\rm A}.8b)  \cr   } $$
This formula gives the leading term $\ln r/r$ and the sub-dominant term  
$1/r$. If we try to find the leading term using (A.7) instead of (A.5) [i.e.  
by changing the variable $s=r+\lambda$ in (A.7) and expanding the integrand
when $r\rightarrow \infty$, $t-r=$~const], we get formally a $1/r$ term but
in factor of a divergent integral, as expected since the leading term
is actually $\ln r/r$.

In the case of a source term corresponding to $k\geq 3$, the relevant formula  
is obtained by performing $k-2$ integrations by parts of (A.1). 
We get in this way

$$ \eqalignno{
  \square^{-1}_R \biggl[ \left( {r\over r_0} \right)^{\!B} {\hat n_L\over r^k}
  F(t-r) \biggr] &= \left({r\over r_0} \right)^{\!B} \sum^{k-3}_{i=0}
  \alpha_i (B) \hat n_L {F^{(i)} (t-r)\over r^{k-i-2}} \cr
  &+ \beta (B) \square^{-1}_R
  \biggl[ \left( {r\over r_0} \right)^{\!B} {\hat n_L\over r^2}
  F^{(k-2)} (t-r) \biggr]  \ , &({\rm A}.9) \cr } $$

\noindent where the second term is a retarded integral of the type
studied before, and where the coefficients are

$$ \eqalignno{
  \alpha_i (B) &= {2^i (B-k+2+i)..(B-k+3)\over
   (B-k+2-\ell +i)..(B-k+2-\ell)(B-k+3+\ell+i)..(B-k+3+\ell)}\ , \cr
 & & ({\rm A}.10a) \cr
  \beta (B) &= {2^{k-2} B(B-1)..(B-\ell)\over
   (B-k+2)..(B-k-\ell+2)(B+\ell)..(B-k+\ell+3)} \ .
 & ({\rm A}.10b) \cr  } $$
The factors symbolized by dots decrease by steps of one unit from left
to right. Before taking the finite part one must study the occurence of  
poles at $B=0$ in the coefficients (A.10). Two cases must be distinguished.  
In the case $3\leq k\leq \ell +2$, none of the denominators in (A.10) vanish  
at $B=0$. This implies that the second term in (A.9) is zero at $B=0$,  
owing to the explicit factor $B$ in the numerator of $\beta (B)$. Thus we  
obtain (in the case $3\leq k\leq l+2$) a finite result when $B \to0$,  
which is local and independent of $r_0$,

$$ \eqalignno{
  \square^{-1}_R \biggl[ \left( {r\over r_0} \right)^B {\hat n_L\over r^k}
  F(t-r) \biggr]_{|_{B=0}} =& - {2^{k-3} (k-3)!(\ell+2-k)!\over (\ell+k-2)!}
 \, \hat n_L  \cr
 &\times \sum^{k-3}_{j=0}  {(\ell+j)!\over 2^j j!(\ell-j)!}
  {F^{(k-3-j)} (t-r)\over r^{j+1}} \ . \qquad\qquad &({\rm A}.11) \cr } $$
This formula can be checked by applying the d'Alembertian operator on
both sides. When $r\to\infty$, $t-r=$~const, it gives

$$ \eqalignno{
  \square^{-1}_R \biggl[ \left( {r\over r_0} \right)^{\!B} {\hat n_L\over r^k}
 F(t-r) \biggr]_{|_{B=0}} =& - {2^{k-3} (k-3)!(\ell+2-k)!\over (\ell+k-2)!}
 \hat n_L {F^{(k-3)} (t-r)\over r} \cr
  & + O\left( {1\over r^2} \right)\ .  & ({\rm A}.12) \cr  } $$

Next, in the case $k\geq \ell+3$, the denominators of both $\alpha (B)$  
and $\beta (B)$ vanish at $B=0$, so $\alpha (B)$ involves a (simple) pole,  
while $\beta (B)$ is finite (because the pole is compensated by the factor  
$B$ in the numerator).
In this case ($k\geq\ell +3$) the finite part reads

$$ \eqalignno{
 &\hbox{FP}_{B=0}\, \square^{-1}_R \biggl[ \left( {r\over r_0} \right)^B
  {\hat n_L\over r^k} F(t-r) \biggr] \cr
 &\qquad\qquad = \sum^{k-3}_{i=0} \tilde\alpha_i \hat n_L
 { F^{(i)} (t-r)\over r^{k-i-2}}
 + {(-)^{k+\ell} 2^{k-2} (k-3)! \over (k+\ell-2)!(k-\ell-3)!}  \cr
 &\qquad\qquad \times \left\{ {(-)^\ell\over 2} \ln \left( {r\over r_0}\right)
  \hat\partial_L \left( {F^{(k-\ell-3)} (t-r)\over r } \right)
  +\square^{-1}_R \biggl[ {\hat n_L\over r^2} F^{(k-2)} (t-r)
   \biggr] \right\}\ . \cr
   &  & ({\rm A}.13) \cr  } $$
The coefficients are given, when $0\leq i\leq k-\ell-4$, by

$$  \tilde\alpha_i = {(-2)^i (k-3)!\over (k+\ell-2)!(k-\ell-3)!}
  {(k+\ell-3-i)!(k-\ell-4-i)!\over (k-3-i)!} \ , \eqno({\rm A}.14a)  $$

\noindent and, when $k-\ell-3\leq i\leq k-3$, by

$$\eqalignno{
  \tilde\alpha_i = & {2^i(-)^{k+\ell} (k-3)!\over (k+\ell-2)!(k-\ell-3)!}
 {(k+\ell-3-i)!\over (k-3-i)! (\ell-k+i+3)!} \cr
 &\times \biggl\{ \sum^{k-\ell-3}_{j=1} {1\over j}
  -\sum^{\ell+3+i-k}_{j=1} {1\over j}
  -\sum^{k-3}_{j=k-2-i} {1\over j} +\sum^{k+\ell-2}_{j=k+\ell-2-i} {1\over j}
  \biggr\} \ . &({\rm A}.14b) \cr } $$
[One can express (A.14) with the help of the Euler $\Gamma$-function and
its logarithmic derivative $\psi$.] Note that (A.13) has a genuine dependence  
on the length scale $r_0$ through the logarithm of $r/r_0$, which is in  
factor of a homogeneous solution, say

$$ \hat\partial_L \left( {G(t-r)\over r} \right) = (-)^\ell \hat n_L
\sum^\ell_{j=0} {(\ell+j)!\over 2^j j!(\ell-j)!}
    {G^{(\ell-j)}(t-r) \over r^{j+1}} \ . \eqno({\rm A}.15) $$
The last term in (A.13) involves a non-local integral known from
(A.3)--(A.5) or (A.7). There exists a special combination of retarded
integrals (A.13) which is purely local, finite at $B=0$, logarithm-free
and independent of $r_0$.  This combination, particularly useful in
practical computations, reads

$$ \eqalignno{
&\square^{-1}_R \biggl[ \left( {r\over r_0} \right)^B \hat n_L \left( 2(k-2)
 {F^{(1)} (t-r)\over r^k} +[(k-1)(k-2) -\ell (\ell+1)]
 {F (t-r)\over r^{k+1}}\right) \biggr]_{|_{B=0}} \cr
&\qquad\qquad = \hat n_L
 {F (t-r)\over r^{k-1}} + \gamma \hat \partial_L  \left(
 {F^{(k-\ell-2)} (t-r)\over r} \right) \ , & ({\rm A}.16a) \cr } $$

\noindent where

$$   \gamma = {(-)^{k}2^{k-2} (k-3)! \over (k+\ell-1)!(k-\ell-2)!}
  \Bigl[ (k+\ell-1)(k-\ell-2)-(2k-3)(k-2) \Bigr] \ . \eqno({\rm A}.16b) $$
The formula is valid when $k\geq \ell+2$; when $k=\ell+2$ we recover
(A.11). The second term in (A.16a) is an homogeneous solution fixed by  
our particular way of integrating the wave equation. Thus the first term  
by itself is a solution of the required equation, but the homogeneous  
solution must absolutly be taken into account when doing practical
computations with this method.

To leading-order when $r\to \infty$ with $t-r=$~const, (A.13) reads

$$ \eqalignno{
  \hbox{FP}_{B=0} \square^{-1}_R&\biggl[ \left( {r\over r_0} \right)^{\!B}
   {\hat n_L\over r^k} F(t-r) \biggr]
 = {(-)^{k+\ell}2^{k-3} (k-3)! \over (k+\ell-2)!(k-\ell-3)!}
   {\hat n_L \over r} \cr
& \times \int^{+\infty}_0 d\lambda F^{(k-2)}
  (t-r -\lambda) \biggl[ \ln \left( {\lambda\over 2r_0}\right)
 +\sum^{k-\ell-3}_{j=1} {1\over j}
 +\sum^{k+\ell-2}_{j=k-2} {1\over j} \biggr] \cr
 &+ O \left( {1\over r^2} \right) \ . & ({\rm A}.17) \cr } $$
As in the case $k=2$ given by (A.8b), the leading $1/r$ term is
non-local. However, in contrast with (A.8b), the expansion is free of
logarithms of $r$ (this is true to all orders in $1/r$), and
depends irreducibly on the length scale $r_0$.

Finally we give the expression of the pole part when
$B\to 0$ of the retarded integral in the case $k\geq \ell+3$. The
result, easily obtained from (A.9)-(A.10), is

$$  \square^{-1}_R \biggl[ B \left( {r\over r_0} \right)^{\!B}
   {\hat n_L\over r^k} F(t-r) \biggr]_{|_{B=0}}
   = {(-)^k 2^{k-3} (k-3)! \over (k+\ell-2)!(k-\ell-3)!} \hat \partial_L
 \left( {F^{(k-\ell-3)} (t-r)\over r} \right)
\ . \eqno({\rm A}.18) $$
As there are only simple poles, the result is zero when the power of
$B$ is strictly larger than one.

\appendix{The quadrupole-(current-)dipole interaction}  

We start from the linearized metric (2.3) written for the mass quadrupole
$M_{ab}$ and the (static) current dipole $S_a$,
$$\eqalignno{
h^{00}_1 =& -2 \partial_{ab} (r^{-1}M_{ab}),&({\rm B}.1a)  \cr
h^{0i}_1 =& -2 \varepsilon_{iab} \partial_a (r^{-1}) S_b + 2\partial_a
(r^{-1}M_{ai}^{(1)}), &({\rm B}.1b) \cr
h^{ij}_1 =& -2r^{-1}M^{(2)}_{ij},& ({\rm B}.1c) \cr }
$$
and replace it into (2.5), keeping only the terms $M_{ab}\times S_c$.

Once the source $N_2^{\mu \nu}$ for this interaction is known in
all-expanded form (3.3), we can apply on each of the terms the formulas
of Appendix A. However, as we are only interested in the two
coefficients of the terms $M_{ab}\times S_c$ in (5.10) and (5.12),
it is sufficient to obtain the $1/r$ part of the metric $h^{\mu \nu}_2$  
when $r \rightarrow \infty$, $t-r=$ const. Thus we proceed like in  
Appendix C of [24], which computed the $1/r$ part of $h^{\mu \nu}_2$  
in the cases of quadrupole-monopole interactions. We recall the necessary  
formulas derived in [24],

$$ \eqalignno{
 &{\rm FP}_{B=0} \square^{-1}_R \left[ r^{B-1}\hat\partial_L \left(
  r^{-1} F (t-r)
 \right)\right] ={(-)^\ell\hat n_L\over 2r} \int^\infty_0 d\lambda F^{(\ell)}
 (t-r-\lambda) \cr
  &\qquad\qquad\qquad\qquad\times\biggl[ \ln \left( {\lambda \over 2r}\right)
 + \sum^\ell_{j=1} {1\over j} \biggr] + O\left( {\ln r\over r^2}\right)\ ,
 &({\rm B}.2a) \cr
 &{\rm FP}_{B=0} \square^{-1}_R\left[ r^B \partial_i (r^{-1}) \hat\partial_L
\left( r^{-1} F (t-r) \right)\right]  \cr
 &\qquad\qquad\qquad
 = {(-)^\ell\over 2(\ell+1)} (n_i\hat n_L -\delta_{i< a_\ell}
n_{L-1>})
 {F^{(\ell)} (t-r)\over r} + O\left( {1\over r^2}\right)
  \ ,  &({\rm B}.2b) \cr
 &{\rm FP}_{B=0} \square^{-1}_R\left[ r^B\partial_{ij} (r^{-1})
\hat\partial_L
  \left( r^{-1} F (t-r) \right)\right] \cr
  &= {(-)^{\ell+1}\over 2(\ell+1)(\ell+2)} \biggl\{ -{4\over 3}
\delta_{ij}\delta_{\ell0}
  +(n_{ij}+\delta_{ij})\hat n_L  \cr
  &- 2 [\delta_{i<a_\ell} n_{L-1>}  n_j
    +\delta_{j<a_\ell} n_{L-1>}n_i]
   + 2\delta_{i<a_\ell}\delta_{\underline{j}a_{\ell-1}} n_{L-2>}
 \biggr\} {F^{(\ell+1)} (t-r)\over r}\cr
   & \qquad\qquad\qquad + O\left( {1\over r^2}\right) \ . &({\rm B}.2c)\cr }
$$
We have added in (B.2c) the term $-{4\over 3} \delta_{ij}  
\delta_{\ell0}$ (where $\delta_{ij}$ and $\delta_{\ell0}$ are Kronecker  
symbols) with respect to the formula (C3) given in Appendix C of [24].  
Indeed this term is mistakenly missing in the formula (C3) of [24] (but it  
does not change any of the results derived in [24]).

The formulas (B.2) yield straightforwardly the $1/r$ term in the first
part $u^{\mu\nu}_2$ of the metric [see (2.8)]. Then we compute the
($1/r$ term of the) divergence $w^\mu_2=\partial_\nu u^{\mu\nu}_2$ and
deduce from (2.11)-(2.12) the second part $v^{\mu\nu}_2$ of the
metric. We find that $v^{\mu \nu}_2$ is non zero in the case $M_{ab}\times
S_c$, contrarily to the cases $M\times M_L$ and $M\times S_L$ studied  
in [24] for which the $v_2^{\mu\nu}$'s are zero. The result of the computation
is

$$\eqalignno{
  h^{00}_2 =& -{4\over 3} r^{-1} \varepsilon_{abc} n_{ad} M^{(4)}_{bd}
    S_c + O(r^{-2})\ , &({\rm B}.3a) \cr
h^{0i}_2 =& r^{-1} \left[ -{4\over 3} \varepsilon_{abc} n_a
M^{(4)}_{ib} S_c
 - {5\over 6} \varepsilon_{iab} n_{acd} M^{(4)}_{cd} S_b \right]
+ O(r^{-2})\ ,
&({\rm B}.3b) \cr
h^{ij}_2 =& r^{-1} \left[ {1\over 3} \varepsilon_{abc} n_{ijad} M^{(4)}_{bd}
  S_c - {8\over 3} \varepsilon_{abc} n_{a(i}M^{(4)}_{j)b} S_c \right.\cr
&+ \delta_{ij} \varepsilon_{abc} n_{ad} M^{(4)}_{bd} S_c - 2
\varepsilon_{ab(i} n_{ac} M^{(4)}_{j)c}S_b  \cr
&\left.+ {1\over 3} \varepsilon_{ab(i} n_{j)acd} M^{(4)}_{cd} S_b
 \right]+ O(r^{-2})\ . &({\rm B}.3c)\cr}
$$
From (B.3c) we obtain the TT projection and then deduce the associated  
observable moments. Only the mass quadrupole and current octupole  
moments receive a contribution,
$$\eqalignno{
\delta U_{ij} =& {1\over 3} \varepsilon_{ab<i} M^{(4)}_{j>a} S_b,
      &({\rm B}.4a)\cr
\delta V_{ijk} =& - 2M^{(4)}_{<ij} S_{k>}.&({\rm B}.4b) \cr}
$$
We thus obtain the coefficients quoted in (5.10) and (5.12) of the text.

\references
\numrefjl{[1]}{Peters P C 1966}{Phys. Rev.}{146}{938}
\numrefjl{[2]}{Price R H 1972}{Phys. Rev. D}{5}{2419}
\numrefjl{[3]}{Bardeen J M and Press W H 1973}{J. Math. Phys.}{14}{7}
\numrefjl{[4]}{Bonnor W B and Rotenberg M A 1966}{Proc. R. Soc.
London A}{289}{247}
\numrefjl{[5]}{Couch W E, Torrence R J, Janis A I and
Newman E T 1968}{J. Math. Phys.}{9}{484}
\numrefjl{[6]}{Hunter A J and Rotenberg M A 1969}{J. Phys. A}{2}{34}
\numrefbk{[7]}{Bonnor W B 1974 in}{Ondes et radiations
gravitationnelles}{(CNRS, Paris) p. 73}
\numrefjl{[8]}{Blanchet L and Damour T 1988}{Phys. Rev. D}{37}{1410}
\numrefjl{[9]}{Blanchet L and Damour T 1992}{Phys. Rev. D}{46}{4304}
\numrefjl{[10]}{Bonnor W B 1959}{Philos. Trans. R. Soc. London A}{251}{233}
\numrefjl{[11]}{Bonnor W B and Rotenberg M A 1961}{Proc. R. Soc.
London A}{265}{109}
\numrefjl{[12]}{Payne P N 1983}{Phys. Rev. D}{28}{1894}
\numrefbk{[13]}{Blanchet, L 1990 Th\`ese d'habilitation, Universit\'e P.
et M. Curie (1990) (unpublished)}{}{}
\numrefjl{[14]}{Christodoulou D 1991}{Phys. Rev. Lett.}{67}{1486}
\numrefjl{[15]}{Thorne K S 1992}{Phys. Rev. D}{45}{520}
\numrefjl{[16]}{Wiseman A G and Will C M 1991}{Phys. Rev. D}{44}{R2945}
\numrefjl{[17]}{Braginsky V B and Grishchuk L P 1985}{Sov. Phys.
JETP}{62}{427}
\numrefjl{[18]}{Braginsky V B and Thorne K S 1987}{Nature}{327}{123}
\numrefbk{[19]}{Blanchet L 1997 in}{Relativistic gravitation and gravitational
radiation}{Lasota J P and Marck J A eds. (Cambridge Univ. Press) p. 33}
\numrefjl{[20]}{Blanchet L and Damour T 1986}{Philos. Trans. R. Soc.
London A}{320}{379}
\numrefjl{[21]}{Blanchet L 1987}{Proc. R. Soc. Lond. A}{409}{383}
\numrefjl{[22]}{Thorne K S 1980}{Rev. Mod. Phys.}{52}{299}
\numrefbk{[23]}{Bonnor W B and Piper M S}{Implosion of quadrupole
gravitational waves}{gr-qc/9610015}
\numrefjl{[24]}{Blanchet L 1995}{Phys. Rev. D}{51}{2559}
\numrefjl{[25]}{Blanchet L 1996}{Phys. Rev. D}{54}{1417}
\numrefjl{[26]}{Blanchet L}{Gravitational-wave tails of tails}{}{(referred
to as paper II)}
\numrefjl{[27]}{Cutler C, Apostolatos T A, Bildsten L, Finn L S,
Flanagan E E, Kennefick D, Markovic D M, Ori A, Poisson E,
Sussman G J and Thorne K S 1993}{Phys. Rev. Lett.}{70}{2984}
\numrefjl{[28]}{Cutler C, Finn L S, Poisson E and Sussmann G J
1993}{Phys. Rev. D}{47}{1511}
\numrefjl{[29]}{Tagoshi H and Nakamura T 1994}{Phys. Rev. D}{49}{4016}
\numrefjl{[30]}{Poisson E 1995}{Phys. Rev. D}{52}{5719}
\numrefjl{[31]}{Our notation is the following:  signature $-+++$;  greek
indices =0,1,2,3;  latin indices =1,2,3;  $g={\rm det}$ $(g_{\mu\nu})$;
$\eta_{\mu\nu}= \eta^{\mu\nu}$ =~flat metric =~diag (-1,1,1,1);
$r=|{\bf x}|=(x_1^2 +x_2^2 +x_3^2)^{1/2}$;  $n^i =n_i =x^i/r$;
$\partial_i =\partial/\partial x^i$; $n^L =n_L= n_{i_1} n_{i_2}\ldots
n_{i_\ell}$ and $\partial_L =\partial_{i_1} \partial_{i_2}\ldots
\partial_{i_\ell}$, where $L=i_1 i_2\ldots i_\ell$ is a multi-index with
$\ell$ indices; $n_{L-1} = n_{i_1} \ldots n_{i_{\ell-1}}$, $n_{aL-1}
=n_a n_{L-1}$, etc\dots; $\hat n_L$ and $\hat\partial_L$ are the (symmetric)
and trace-free (STF) parts of $n_L$ and $\partial_L$, also denoted by
$n_{<L>}$, $\partial_{<L>}$;  the superscript $(n)$ denotes $n$ time
derivatives;  $T_{(\alpha\beta)} ={1\over 2} (T_{\alpha\beta} +
T_{\beta\alpha})$ and $T_{(ij)} = {1\over 2} (T_{ij} + T_{ji})$}{}{}{}
\numrefbk{[32]}{Wolfram S 1992}{Mathematica}{second edition
(Addison-Wesley Publishers)}
\numrefbk{[33]}{Fock V A 1959}{Theory of Space, Time and
Gravitation}{Pergamon, London}
\numrefjl{[34]}{Madore J  1970}{Ann. Inst. H. Poincar\'e}{12}{285 and 365}
\numrefbk{[35]}{Whittaker E T and Watson G N 1990}{A course of Modern
Analysis}{reprinted edition (Cambridge Univ. Press)}

\bye